\documentstyle[12pt,subeqnarray,psfig]{article}
\begin{document}
%
%
%
\newenvironment{lefteqnarray}{\arraycolsep=0pt\begin{eqnarray}}
{\end{eqnarray}\protect\aftergroup\ignorespaces}
\newenvironment{lefteqnarray*}{\arraycolsep=0pt\begin{eqnarray*}}
{\end{eqnarray*}\protect\aftergroup\ignorespaces}
\newenvironment{leftsubeqnarray}{\arraycolsep=0pt\begin{subeqnarray}}
{\end{subeqnarray}\protect\aftergroup\ignorespaces}
\newcommand{\appleq}{\stackrel{<}{\sim}}
\newcommand{\appgeq}{\stackrel{>}{\sim}}
\newcommand{\arcsinh}{\mathop{\rm arcsinh}\nolimits}
\newcommand{\arctg}{\mathop{\rm arctg}\nolimits}
\newcommand{\diff}{{\rm\,d}}
\newcommand{\displayfrac}[2]{\frac{\displaystyle #1}{\displaystyle #2}}
\newcommand{\Erfc}{\mathop{\rm Erfc}\nolimits}
\newcommand{\Int}{\mathop{\rm Int}\nolimits}
\newcommand{\Nint}{\mathop{\rm Nint}\nolimits}
\newcommand{\pprime}{{\prime\prime}}

\title{A numerical fit of analytical to simulated 
       density profiles 
       in dark matter haloes}
\author{{R. Caimmi, C. Marmo, T. Valentinuzzi\footnote{
{\it Astronomy Department, Padua Univ., Vicolo Osservatorio 2,
I-35122 Padova, Italy}
email: caimmi@pd.astro.it}
\phantom{agga}}}





\maketitle
\begin{quotation}
\section*{}
\begin{Large}
\begin{center}

Abstract

\end{center}
\end{Large}
\begin{small}


Analytical and geometrical properties
of generalized power-law (GPL) density profiles
are investigated in detail.   In particular, a
one-to-one correspondence is found between 
mathematical parameters (a scaling radius, $r_0$,
a scaling density, $\rho_0$, and three exponents,
$\alpha$, $\beta$, $\gamma$), and geometrical 
parameters (the coordinates of the intersection 
of the asymptotes, $x_C$, $y_C$, and three 
vertical intercepts, $b$, $b_\beta$, $b_\gamma$,
related to the curve and the asymptotes,
respectively): $(r_0,\rho_0,\alpha,\beta,\gamma)
\leftrightarrow(x_C,y_C,b,b_\beta,b_\gamma)$.
Then GPL density profiles are compared with
simulated dark haloes (SDH) density profiles,
and nonlinear least-absolute values and
least-squares fits involving the above mentioned
five parameters (RFSM5 method) are prescribed.
More specifically, the sum of absolute values or 
squares of absolute logarithmic residuals, $R_i=
\log\rho_{SDH}(r_i)-\log\rho_{GPL}(r_i)$, is
evaluated on $10^5$ points making a 5-dimension
hypergrid, through a few iterations. The
size is progressively reduced around a fiducial
minimum, and superpositions on nodes of earlier
hypergrids are avoided.   An application is
made to a sample of 17 SDHs on the scale of
cluster of galaxies, within a flat $\Lambda$CDM
cosmological model (Rasia et al. 2004).   In dealing
with the mean SDH density profile, a virial
radius, $r_{{\rm vir}}$, averaged over the whole
sample, is assigned, which allows the calculation
of the remaining parameters.   Using a RFSM5
method provides a better fit with respect to
other methods.   The geometrical parameters,
averaged over the whole sample of best fitting
GPL density profiles, yield $(\alpha,\beta,
\gamma)\approx(0.6,3.1,1.0)$, to be compared
with $(\alpha,\beta,\gamma)=(1,3,1)$, i.e. the
NFW density profile (Navarro et al. 1995, 1996,
1997); $(\alpha,\beta,\gamma)=(1.5,3,1.5)$
(Moore et al. 1998, 1999); $(\alpha,\beta,
\gamma)=(1,2.5,1)$ (Rasia et al. 2004); and,
in addition, $\gamma\approx1.5$ (Hiotelis 2003),
deduced from the application of a RFSM5 method,
but using a different definition of scaled
radius, or concentration; and $\gamma\approx$
1.2\,-1.3 deduced from more recent high-resolution
simulations (Diemand et al. 2004; Reed et al.
2005).   No evident
correlation is found between SDH 
dynamical state (relaxed or merging) and
asymptotic inner slope of the fitting logarithmic
density profile or (for SDH comparable
virial masses) scaled radius.  Mean values and standard
deviations of some parameters are calculated,
and in particular the decimal logarithm of the
scaled radius, $\xi_{{\rm vir}}$,
reads $<\log\xi_{{\rm vir}}>=0.74$ and $\sigma_{s
\log\xi_{{\rm vir}}}=0.15$-0.17, consistent with previous
results related to NFW density profiles.   It
provides additional support to the idea, that
NFW density profiles may be considered as a
convenient way to parametrize SDH density
profiles, without implying that it necessarily
produces the best possible fit (Bullock et al. 
2001).   A certain degree of degeneracy is
found in fitting GPL to SDH density profiles.
If it is intrinsic to the RFSM5 method or it 
could be reduced by the next generation of
high-resolution simulations, still remains
an open question. 

\noindent
{\it keywords - cosmology: theory -- dark matter -- galaxies:
clusters -- galaxies: haloes.}


\end{small}
\end{quotation}

%

\section{Introduction}\label{intro}
Recent observations of anisotropies in the cosmic
microwave background put severe constraints on the 
cosmological parameters, by comparison with the
predictions of current theories of structure
formation and the evolution of the Universe.   The
addition of information from large-scale structure
surveys, Hubble parameter determinations, and Type
Ia supernova results, yields evidence for a flat
$(\Omega_m+\Omega_\Lambda\approx1)$, low-density
$(\Omega_m\approx0.3;~\Omega_\Lambda\approx0.7)$
universe, a Zeldovich power-law index of primordial 
fluctuations $(n_s\approx1)$, a (non baryonic)
dark matter
density $(\Omega_dh^2\approx0.16)$ dominant
over baryon matter density $(\Omega_bh^2\approx
0.023)$, a Hubble parameter (normalized to 100
km s$^{-1}$Mpc$^{-1}$) between one half and unity 
$(h\approx0.69)$, and a cosmological age between
ten and fifteen billion years $(T\approx13.7$ Gyr), 
which make the main information (Sievers et al. 
2003).   The above mentioned values are consistent
with the results deduced by different investigations
(e.g., Rubi$\tilde{{\rm n}}$o-Martin et al. 2003; Spergel 
et al. 2003; Fukugita \& Peebles 2004).

The related ($\Lambda$CDM) cosmological model (e.g.,
Diemand et al. 2004; Reed et al. 2005) is
consistent with a bottom-up picture (hierarchical
clustering) of dark matter haloes, where smaller
systems formed first from initial density 
perturbations and then merged with each other to
become larger systems, or were tidally disrupted
and accreted from previously formed larger systems.
Most naturally, the density profiles of haloes are
expected to be a two-parameters family.   This is
why, assuming that halo formation may be approximated
to an acceptable extent by spherical collapse, each
protohalo perturbation is characterized by two 
independent parameters e.g., mass and radius (or
overdensity), at some fiducial cosmological time.

A successful two-parameter functional form for the
halo profile, where a scaled density depends on a
scaled radius, has been proposed by Navarro et al.
(1995, 1996, 1997, the last quoted hereafter as 
NFW).   They also argued that the density profile
is universal, in the sense that its shape does not
appreciably change (over two decades in radius)
for different halo masses (spanning about four 
orders of magnitude), initial density fluctuation
spectra, or cosmological parameters.   Many studies
on the NFW ``universal density profile'', both
numerical (with increasing resolution) and 
analytical, were done after their proposal (for 
a more detailed discussion see e.g., Hess et al.
1999; Klypin et al. 2001; Bullock et al. 2001;
Fukushige \& Makino 2001, 2003; M\"uchet \& Hoeft
2003; Zhao et al. 2003).

Generally speaking, the NFW density profile may
be considered as a special case of the 5-parameter
family (Hernquist 1990; Zhao 1996, 1997):
\begin{equation}
\label{eq:GPL}
\rho\left(\frac r{r_0}\right)=\frac{\rho_0}{(r/r_0)^\gamma
[1+(r/r_0)^\alpha]^\chi}~~;\quad\chi=\frac{\beta-\gamma}
\alpha~~;
\end{equation}
related to the choice $(\alpha,\beta,\gamma)=(1,3,1)$,
where $\rho_0$ and $r_0$ are a scaling density and a
scaling radius, respectively
\footnote{To the knowledge of the authors, the
family of density profiles expressed by Eq.\,(\ref
{eq:GPL}) was first defined by Hernquist
[1990, Eq.\,(43) therein], even if his attention
was restricted to  the special case, $(\alpha,\beta,
\gamma)=(1,4,1)$ which, in turn, was earlier proposed
by Kuzmin \& Veltmann (1973).   A family of density
profiles including the special case studied by Hernquist,
the so called $\gamma$ models, where $(\alpha,\beta,
\gamma)=(1,4,\gamma)$, was given independently
by Dehnen (1993) and Tremaine et al. (1994).   A
detailed investigation of general $(\alpha,\beta,\gamma)$
models was performed by Zhao (1996, 1997).}.

The density profile, expressed by Eq.\,(\ref{eq:GPL}),
reduces to a power-law both towards the centre, $r\to0$,
and towards infinite, $r\to+\infty$, where the exponent
equals $-\gamma$ and $-\beta$, respectively.
It may be conceived as a generalized power-law and, in the
following, it shall be quoted as GPL density profile.
On the other hand, matter distribution within a
simulated dark matter halo, in the following, shall be
quoted as SDH density profile.

For fixed exponents, one among the two remaining
free parameters, the scaling density and the scaling
radius, may be related to the mass and the radius of
the virialized region.   For further details see e.g.,
NFW; Bullock et al. (2001); Rasia et al. (2004). 

Some doubts on the ``universality'' of the NFW
density profile were cast by latter investigations.
There are main orders of reasons against the idea
of a universal, NFW density profile, which can be
briefly summarized as follows.
\begin{description}
\item[\rm{(i)}] A steeper slope in the central
regions (e.g., Fukushige \& Makino 1997, 2001,
2003; Moore et al. 1998, 1999; Ghigna et al.
2000; but see Navarro et al. 2004, for a different
point of view).
\item[\rm{(ii)}] A non universal slope in the
central regions, which depends on the power spectrum
of the initial density perturbation (Syer \&
White 1998), or on the mass (Jing \& Suto 2000;
Ricotti 2003).  Additional support to this idea
is provided by recent, high-resolution
simulations (e.g., Fukushige et al. 2004; Navarro
et al. 2004).
\item[\rm{(iii)}] A certain degree of degeneracy
with regard to the exponents, $(\alpha,\beta,
\gamma)$, in fitting various GPL to SDH density
profiles, in the whole range of resolved scales 
(e.g., Klypin et al. 2001).
\item[\rm{(iv)}] Different criterions in
fitting GPL to SDH density profiles, such as
minimizing the maximum fractional deviations
of the fit, $\max\vert\log(\rho_{GPL}/\rho_h)-
\log(\rho_{SDH}/\rho_h)\vert$ (e.g., Klypin et al.
2001); the sum of the squares of absolute%
\footnote{The term ``absolute'' here has not
to be intended as ``absolute value'', but as 
opposite to ``relative''.   More precisely,
$y_i-y(x_i)$ is an absolute residual, while
$[y_i-y(x_i)]/y(x_i)$ is the corresponding
relative residual.}%
logarithmic
residuals, $\chi^2=\sum[\log(\rho_{GPL}/\rho_h)-
\log(\rho_{SDH}/\rho_h)]^2$ (e.g., Bullock et al.
2001); the sum of squares of relative residuals,
$\sum[(\rho_{SDH}/\rho_h-\rho_{GPL}/\rho_h)/(\rho_
{GPL}/\rho_h)]^2$ (e.g., Fukushige \& Makino 2003,
2004), where $\rho_h$ is a normalization value.
For a more detailed discussion see e.g.,
Tasitsiomi et al. (2004).
\item[\rm{(v)}] A gentler slope in the central
regions $(\gamma<1)$ and a steeper slope
sufficiently far from the centre $(\beta>3)$,
under reasonable boundary conditions such as
a finite halo mass and force-free halo centre,
and a vanishing density at infinite distance
(M\"ucket \& Hoeft 2003).   Additional support
to this idea is provided by recent high-resolution
simulations (Navarro et al. 2004).
\item[\rm{(vi)}] A gentler slope in the central 
regions $(\gamma\appleq1)$ to be consistent
with rotation curves, deduced from the observations,
of low surface brightness galaxies (McGaugh \&
de Blok 1988; de Blok et al. 2001), the Galaxy
(Binney \& Evans 2001), and dwarf galaxies (van
den Bosh \& Swaters 2001).   For a more detailed
discussion, see M\"ucket \& Hoeft (2003).   In
addition, the validity of the Jeans equation
implies $1\le\gamma\le3$ for dark matter haloes
(Hansen 2004), unless the effects of the baryonic
component are taken into consideration (El-Zant
et al. 2004; Hansen 2004).
\item[\rm{(vii)}] A discrepancy with the dark
matter distribution required to ensure hydrostatic
equilibrium of gas, deduced from measured X-ray
brightness profiles, in clusters of galaxies
(Arieli \& Rephaeli 2003).
\end{description}

Though the resolution of numerical simulations
is increasingly high, still there is no general 
consensus, or insufficient investigation, about
some questions concerning dark matter halo density
profiles, namely: (1) definition and formulation
of universal density profiles (e.g., Huss et al.
1999; Bullock et al. 2001); (2) connection between
GPL and SDH density profiles; (3) dependence of
GPL density profiles on the three exponents,
$(\alpha,\beta,\gamma)$ and the two scaling
parameters, $(r_0,\rho_0)$; (4) extent to
which two or more GPL density profiles fit the
results of numerical simulations (e.g., Klypin
et al. 2001; Fukushige \& Makino 2001, 2003);
(5) degree of degeneration of the three exponents
in fitting GPL to SDH density profiles (e.g.,
Klypin et al. 2001).

The hierarchical collapse of dark matter into
virialized haloes is likely to have played a
key role in the formation of large-scale objects,
such as galaxies and clusters of galaxies.   The
halo profile has a direct dynamical role in
determining the observable parameters of the
baryonic subsystems.   Therefore further investigation
on the above raised questions appears to be 
important.

To this aim, in fitting GPL to SDH
density profiles, both a nonlinear least-absolute
values and a nonlinear least-squares method
are used in the current paper.
The related boundary condition is that both the 
mass and the radius of the virialized region are
determined by the computer outputs and the choice 
of the cosmological parameters.   The main features
of GPL and SDH density profiles are outlined in
sections \ref{GPL} and \ref{SDH}, respectively.
A comparison between GPL and SDH density profiles
is performed in section \ref{GSc}.   Nonlinear
least-absolute values and least-squares fits are
outlined in section \ref{res}.   The subject of 
section \ref{ara} is an application to a sample
of 17 SDHs and the related mean density profile,
on the scale of clusters of galaxies, taken from
Rasia et al. (2004).   The results are then discussed.
Some concluding remarks are drawn in section 
\ref{conc}.   Further investigation on a few 
special arguments is made in the Appendix.

\section{GPL density profiles}\label{GPL}

Plotting GPL density profiles on a logarithmic
plane, $({\sf O}~\log\xi~\log f)$, necessarily
implies use of dimensionless coordinates, 
defined as:
\begin{equation}
\label{eq:csif}
\xi=\frac r{r_0}~~;\qquad f(\xi)=\frac\rho{\rho_0}~~;
\end{equation}
where the scaled radius, $\xi$, can be related 
to density profiles where the isopycnic surfaces
i.e. constant density, are similar and similarly 
placed ellipsoids.   For further details, see
Caimmi \& Marmo (2003).   Accordingly, Eq.\,(\ref
{eq:GPL}) reduces to:
\begin{equation}
\label{eq:SGPL}
f(\xi)=\frac1{\xi^\gamma(1+\xi^\alpha)^\chi}~~;\qquad 
\chi=\frac{\beta-\gamma}\alpha~~;
\end{equation}
independent of the scaling parameters.

The special choice:
\begin{equation}
\label{eq:rhorc0}
\rho^\dagger=2^{\chi}\rho_0~~;\qquad 
r^\dagger=r_0~~;
\end{equation}
translates Eq.\,(\ref{eq:SGPL}) into:
\begin{equation}
\label{eq:csifc}
f(\xi)=\frac{2^{\chi}}{\xi^\gamma(1+\xi^\alpha)^
\chi}~~;\qquad \chi=\frac{\beta-\gamma}\alpha~~;
\end{equation}
which has an immediate interpretation, as:
\begin{equation}
\label{eq:rhorc}
f(1)=1~~;\qquad\rho(r^\dagger)=\rho^\dagger~~;
\end{equation}
the scaling density, $\rho^\dagger$, coincides
with the density on an isopycnic surface,
where the radius equals the scaling radius, $r
=r^\dagger$.   For further details, see Caimmi \&
Marmo (2003).

Scaled GPL density profiles, expressed by 
Eq.\,(\ref{eq:SGPL}), include $\infty^2$
GPL density profiles, expressed by Eq.\,(\ref
{eq:GPL}), for the whole, allowed set of
scaling parameters, $(r_0,\rho_0)$.   A
similar situation occurs for polytropes
(e.g., Caimmi 1980).

As Eqs.\,(\ref{eq:csif}) and (\ref{eq:SGPL})
imply null density at infinite radius, the
mass distribution has necessarily to be 
ended at an assigned isopycnic surface,
which defines a truncation radius, $R$.
The mass within the truncation isopycnic
surface is (Caimmi \& Marmo 2003):
\begin{equation}
\label{eq:Mt}
M=M(R)=\frac{4\pi}3r_0^3\rho_0\nu_{mas}=M_0\nu_{mas}~~;
\end{equation}
where $M_0$ is a scaling mass and the profile factor,
$\nu_{mas}$, has the explicit expression:
\begin{equation}
\label{eq:num}
\nu_{mas}=3\int_0^\Xi f(\xi)\xi^2\diff\xi~~;
\end{equation}
and the integration is carried up to:
\begin{equation}
\label{eq:Csi}
\Xi=\frac R{r_0}~~;
\end{equation}
which may be conceived as a scaled, truncation
radius.

The logarithmic GPL density profile, deduced
from Eq.\,(\ref{eq:SGPL}), is:
\begin{equation}
\label{eq:logf}
\log f=-\gamma\log\xi-\chi\log(1+\xi^\alpha)~~;
\end{equation}
it can be seen that the first and the third
logarithmic derivative, calculated at $\log
\xi=0$ i.e. $r=r_0$, yields:
\begin{lefteqnarray}
\label{eq:dlogf1}
&& \left(\frac{\diff\log f}{\diff\log\xi}\right)_
{\log\xi=0}=-\frac12(\gamma+\beta)~~; \\
\label{eq:dlogf3}
&& \left(\frac{\diff^3\log f}{\diff(\log\xi)^3}\right)_
{\log\xi=0}=0~~; 
\end{lefteqnarray}
which discloses the geometrical meaning of the scaling
radius. 
\begin{trivlist}
\item[\hspace\labelsep{\bf Geometrical meaning of the
scaling radius in GPL density profiles}] \sl
With regard to logarithmic GPL density profiles, the
maximum slope variation rate occurs at the scaling radius,
$\log\xi=\log(r/r_0)=0$, where the slope equals the mean
slope of the related asymptotes, $-\gamma$ and $-\beta$, 
respectively.
\end{trivlist}
In the special case of NFW density profiles, $\gamma=1$,
$\beta=3$, and the slope at the scaling radius equals 
$-2$ (e.g., Bullock et al. 2001; Klypin et al. 2001;
Hiotelis 2003).   For a more detailed discussion, see
Caimmi \& Marmo (2003). 

In the limit of negligible values of the scaled radius,
$\xi$, with respect to unity, Eq.\,(\ref{eq:logf})
reduces to:
\begin{equation}
\label{eq:as0}
\log f=-\gamma\log\xi~~;\qquad\xi\ll1~~;
\end{equation}
which represents, in the logarithmic plane $({\sf O}
\log\xi~\log f)$, a straight line with slope equal to
$-\gamma$ and intercept equal to 0.

In the limit of preponderant values of the scaled radius,
$\xi$, with respect to unity, Eq.\,(\ref{eq:logf})
reduces to:
\begin{equation}
\label{eq:asi}
\log f=-\beta\log\xi~~;\qquad\xi\gg1~~;
\end{equation}
which represents, in the logarithmic plane $({\sf O}
\log\xi~\log f)$, a straight line with slope equal to
$-\beta$ and intercept equal to 0.
 
\begin{figure}
\centerline{\psfig{file=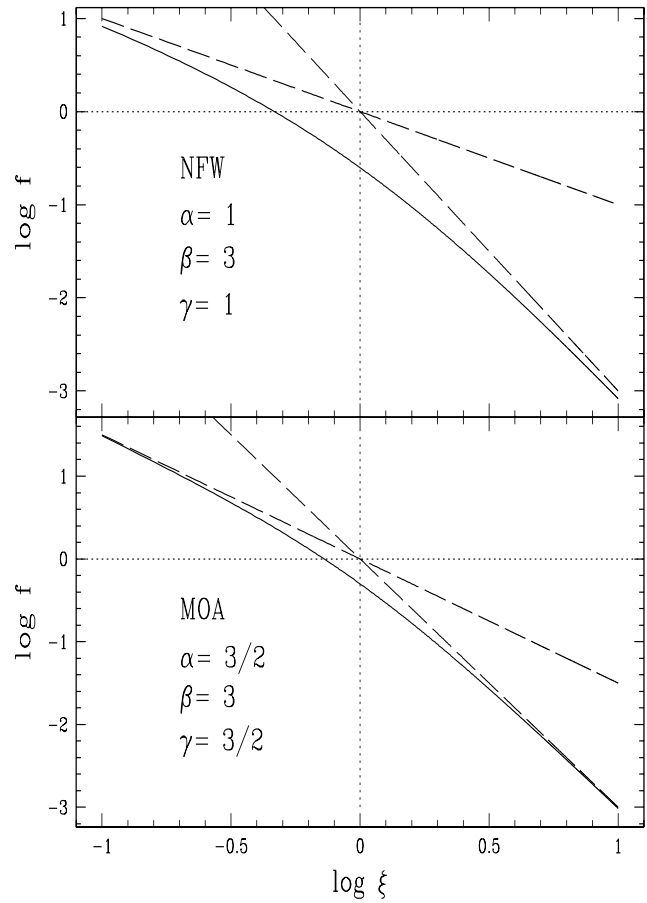,height=130mm,width=90mm}}
\caption{Logarithmic NFW (top) and MOA (bottom)
density profiles (full curves), with their asymptotes
(dashed lines), in the plane $({\sf O}\log\xi~\log 
f)$.
}
\label{f:GPL}    
\end{figure}

It can easily be seen that the straight lines,
expressed by Eqs.\,(\ref{eq:as0}) and (\ref{eq:asi}),
meet at the origin and, in addition, represent the
asymptotes of the logarithmic GPL density profile,
expressed by Eq.\,(\ref{eq:logf}).   The special
cases related to NFW and MOA (Moore et al. 1999)
logarithmic density profiles, are plotted in
Fig.\,\ref{f:GPL}.
The above results hold for $\alpha>0$.   The case
$\alpha<0$ makes the asymptotes change one into
the other.   The limiting case $\alpha=0$ makes
either a vanishing density $(\beta\ne\gamma)$ or
the asymptotes coincide i.e. the curve reduces to
a straight line $(\beta=\gamma)$.

\section{SDH density profiles}\label{SDH}

Dark matter haloes simulations need three basic
ingredients, namely: (i) a cosmological model
with fixed values of the parameters; (ii) an
environment with defined initial conditions;
and (iii) an assigned computer code.   The
density profile during the evolution, is
calculated through the following steps: 
(1) determine the centre of mass of the halo;
(2) count the particles (bound to the halo)
within spherical shells, centered on the 
centre of mass, and equally spaced in
logarithmic distance i.e. $\log(r_{i+1}/
r_i)=$const; (3) evaluate the mean density
within each shell.   For further details
see e.g., NFW; Klypin et al. (2001);
Bullock et al. (2001); Fukushige \&
Makino (2001, 2003).

Simulated haloes are characterized by a
``virial'' parameter, either the virial
mass, $M_{{\rm vir}}$, or the virial radius,
$r_{{\rm vir}}$, defined such that the mean
density inside the virial radius is
$\Delta_{{\rm vir}}$ times the mean matter
universal density, $\rho_h=\rho_{crit}
\Omega_m$, at that redshift:
\begin{equation}
\label{eq:Mvir}
M_{{\rm vir}}=\frac{4\pi}3\Delta_{{\rm vir}}\rho_
{crit}\Omega_mr_{{\rm vir}}^3~~;
\end{equation}
where $\rho_{crit}=3H^2/(8\pi G)$ is
the critical density for closure of
the Universe.
The critical overdensity at virialization,
$\Delta_{{\rm vir}}$, is motivated by the
spherical collapse model: it is below
two hundreds for an Einstein-de Sitter
cosmology, and exceeds three hundreds
for a flat $\Lambda$CDM cosmology where
$\Omega_m\approx0.3$, at $z=0$ (e.g.,
Bullock et al. 2001).

Plotting SDH density profiles on a 
logarithmic plane, $({\sf O}\log\eta~
\log\psi)$, necessarily implies use of
dimensionless coordinates, defined as:
\begin{equation}
\label{eq:SSDH}
\eta=\frac r{r_{{\rm vir}}}~~;\qquad\psi(\eta)=
\frac\rho{\rho_h}~~;\qquad\rho_h=\rho_
{crit}\Omega_m~~;
\end{equation}
without loss of generality.

An upper limit to the domain of SDH 
density profiles follows from the 
definition of virial radius: regions
placed outside are still falling in,
and their macroscopic kinetic energy
has still to be converted into peculiar
energy (e.g., Cole \& Lacey 1996; NFW).

A lower limit to the domain of SDH
density profiles is put by the
occurrence of numerical artifacts
(mainly two-body relaxation) in the
central regions, within about $0.01
r_{{\rm vir}}$ (e.g., Bullock et al. 2001;
Fukushige \& Makino 2001, 2003, 2004;
Navarro et al. 2004) or even less
(e.g., Diemand et al. 2004; Reed et al.
2005).

\begin{figure}
\centerline{\psfig{file=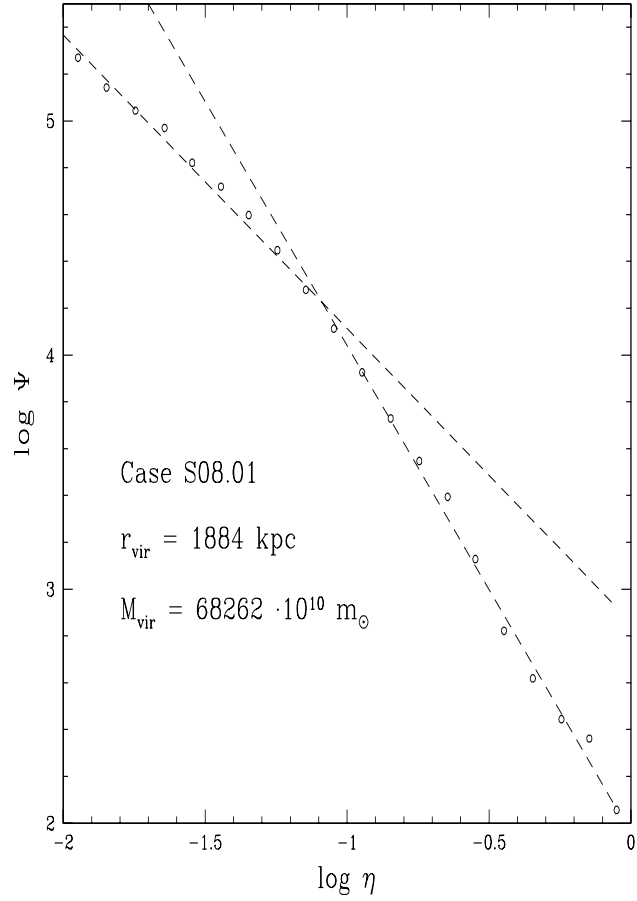,height=130mm,width=90mm}}
\caption{Logarithmic SDH density profile (open circles)
for a typical dark matter halo on the scale of clusters
of galaxies.   The virialized region is safely 
represented in the range $-2<\log\eta<0$.
Also plotted are the best linear fits (dashed),
determined by use of a least-squares fit to
simulated data, within the $\gamma$-region,
$-2<\log\eta<-1$, and the $\beta$-region,
$-1<\log\eta<0$, respectively.}
\label{f:SDH}
\end{figure}

Accordingly, SDH density profiles
appear to be closely related to the
virialized region in the range:
\begin{equation}
\label{eq:doeta}
-2<\log\eta<0~~;
\end{equation}
which shall be assumed in the following.
A typical SDH density profile on the
scale of cluster of galaxies, taken
from a sample of 17 simulated haloes
(Rasia et al. 2004), is represented in
Fig.\,\ref{f:SDH}.
Also plotted therein are the best linear fits,
determined by use of a least-squares fit to
simulated data, within the $\gamma$-region,
$-2<\log\eta<-1$, and the $\beta$-region,
$-1<\log\eta<0$, respectively.   For further
details, see Appendix A.

\section{Comparison between GPL and SDH
density profiles}\label{GSc}

A comparison between GPL and SDH density
profiles necessarily implies that the
truncation radius, and the mass enclosed
within the truncation isopycnic surface,
do coincide with the virial radius and
the virial mass, $R=r_{{\rm vir}}$ and $M=M_
{vir}$, respectively.   Then the combination
of Eqs.\,(\ref{eq:Mt}) and (\ref{eq:Mvir}) 
yields:
\begin{lefteqnarray}
\label{eq:rho0h}
&& \frac{\rho_0}{\rho_h}=\frac{\Delta_{{\rm vir}}\xi_{{\rm vir}}^3}
{\nu_{mas}}~~; \\
\label{eq:csiv}
&& \xi_{{\rm vir}}=\frac{r_{{\rm vir}}}{r_0}~~;
\end{lefteqnarray}
where the scaled virial radius, $\xi_{{\rm vir}}$, is
usually defined as concentration in the special 
case of NFW density profiles (NFW).   With regard
to a generic GPL density profile, there are
several definitions of concentration (e.g.,
Klypin et al. 2001).   Throughout this paper we
shall define the concentration as the scaled
virial radius i.e. the ratio of the virial
radius to the radius where the logarithmic
slope of the density profile equals the mean
slope of the two asymptotes, and the slope
variation rate is maximum, according to Eqs.\,(\ref
{eq:dlogf1}) and (\ref{eq:dlogf3}), respectively. 

The comparison of scaled GPL density profiles,
expressed by Eqs.\,(\ref{eq:csif}), with scaled
SDH density profiles, expressed by Eqs.\,(\ref
{eq:SSDH}), yields:
\begin{equation}
\label{eq:fpsi}
\xi=\xi_{{\rm vir}}\eta~~;\qquad f=\frac{\rho_h}
{\rho_0}\psi~~;
\end{equation}
where $\xi_{{\rm vir}}$ is defined by Eq.\,(\ref
{eq:csiv}).   Accordingly, a generic, scaled
GPL density profile, expressed by Eq.\,(\ref
{eq:SGPL}), takes the equivalent form:
\begin{equation}
\label{eq:PGPL}
\psi(\eta)=\frac{\rho_0/\rho_h}{(\xi_{{\rm vir}}\eta)^
\gamma\left[1+(\xi_{{\rm vir}}\eta)^\alpha\right]^
\chi}~~;\qquad \chi=\frac{\beta-\gamma}\alpha~~;
\end{equation}
and the related, logarithmic GPL density
profile, is deduced by use of Eq.\,(\ref
{eq:rho0h}).  The result is:
\begin{lefteqnarray}
\label{eq:lpsi}
&& \log\psi=\log\Delta_{{\rm vir}}-\log\nu_{mas}+
3\log\xi_{{\rm vir}} \nonumber \\
&& -\gamma\log\xi_{{\rm vir}}-\gamma\log\eta-\chi
\log\left[1+(\xi_{{\rm vir}}\eta)^\alpha\right]~~;
\end{lefteqnarray}
which depends on three exponents, $(\alpha,
\beta,\gamma)$, and two scaling parameters,
$(r_0,\rho_0)$.   On the other hand, the
scaled mass, $\nu_{mas}$, is defined by
Eqs.\,(\ref{eq:num}) and (\ref{eq:Csi});
the virial radius,
$r_{{\rm vir}}$, is known from the computer run;
and the critical overdensity, $\Delta_{{\rm vir}}$,
is determined by the cosmological model.

According to Eqs.\,(\ref{eq:csif}), (\ref
{eq:dlogf1}), (\ref{eq:dlogf3}), and (\ref
{eq:fpsi}), the maximum variation in slope
occurs at $r=r_0$ i.e. $\log\xi=0$ i.e.
$\log\eta=-\log\xi_{{\rm vir}}$.   Then 
Eqs.\,(\ref{eq:dlogf1}) and (\ref{eq:dlogf3})
translate into:
\begin{lefteqnarray}
\label{eq:dlogp1}
&& \left(\frac{\diff\log\psi}{\diff\log\eta}\right)_
{\log\eta=-\log\xi_{{\rm vir}}}=-\frac12(\gamma+\beta)~~; \\
\label{eq:dlogp3}
&& \left(\frac{\diff^3\log\psi}{\diff(\log\eta)^3}\right)_
{\log\eta=-\log\xi_{{\rm vir}}}=0~~; 
\end{lefteqnarray}
where the slope at $\log\eta=-\log\xi_{{\rm vir}}$ equals 
the mean slope of the related asymptotes, $-\gamma$ and 
$-\beta$, respectively.

In the limit of negligible values of the scaled radius,
$\xi_{{\rm vir}}\eta$, with respect to unity, Eq.\,(\ref
{eq:lpsi}) reduces to:
\begin{lefteqnarray}
\label{eq:pas0}
&& \log\psi=\log\Delta_{{\rm vir}}-\log\nu_{mas}+
3\log\xi_{{\rm vir}} \nonumber \\
&& -\gamma\log\xi_{{\rm vir}}-\gamma\log\eta~~;\qquad
\xi_{{\rm vir}}\eta\ll1~~;
\end{lefteqnarray}
which represents, in the logarithmic plane $({\sf O}
\log\eta~\log\psi)$, a straight line with slope equal to
$-\gamma$ and intercept equal to $\log\Delta_{{\rm vir}}-
\log\nu_{mas}+(3-\gamma)\log\xi_{{\rm vir}}$.

In the limit of preponderant values of the scaled radius,
$\xi_{{\rm vir}}\eta$, with respect to unity, Eq.\,(\ref
{eq:lpsi}) reduces to:
\begin{lefteqnarray}
\label{eq:pasi}
&& \log\psi=\log\Delta_{{\rm vir}}-\log\nu_{mas}+
3\log\xi_{{\rm vir}} \nonumber \\
&& -\beta\log\xi_{{\rm vir}}-\beta\log\eta~~;\qquad
\xi_{{\rm vir}}\eta\gg1~~;
\end{lefteqnarray}
which represents, in the logarithmic plane $({\sf O}
\log\eta~\log\psi)$, a straight line with slope equal to
$-\beta$ and intercept equal to $\log\Delta_{{\rm vir}}-
\log\nu_{mas}+(3-\beta)\log\xi_{{\rm vir}}$.
 
\begin{figure}
\centerline{\psfig{file=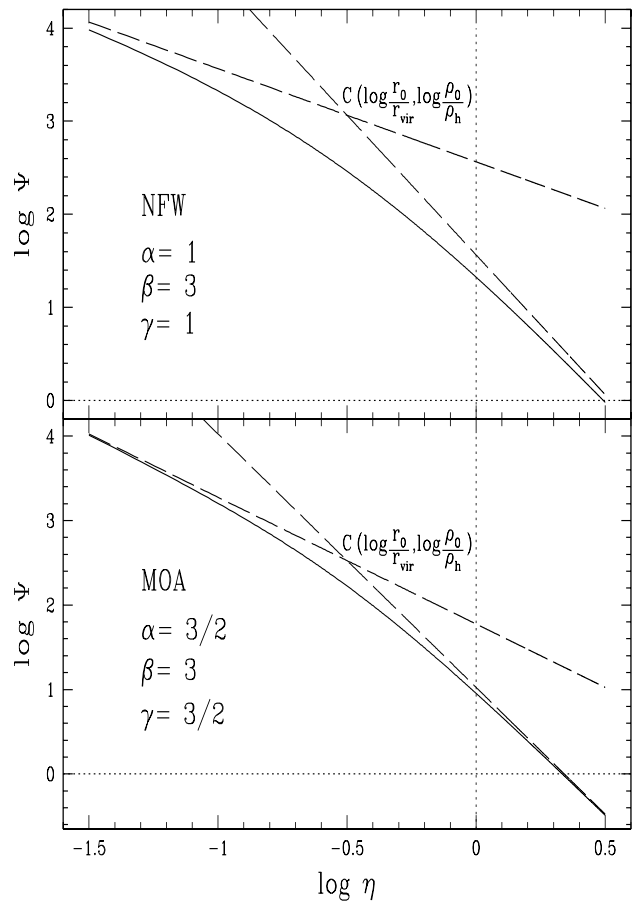,height=130mm,width=90mm}}
\caption{Logarithmic NFW (top) and MOA (bottom) 
density profiles (full curves), with their asymptotes
(dashed lines), in the plane $({\sf O}\log\eta~\log 
\psi)$.   Values of $r_0/r_{{\rm vir}}$ and $\rho_0/\rho_h$
have been arbitrarily chosen.
}
\label{f:GPLs}
\end{figure}

It can easily be seen that the straight lines,
expressed by Eqs.\,(\ref{eq:pas0}) and (\ref
{eq:pasi}), meet at the point $(\log\eta,\log\psi)
=[\log(r_0/r_{{\rm vir}}),\log(\rho_0/\rho_h)]$, where 
Eqs.\,(\ref{eq:rho0h}) and (\ref{eq:csif}) have
been used.  In addition, the above mentioned
straight lines represent the
asymptotes of the logarithmic GPL density profile,
expressed by Eq.\,(\ref{eq:lpsi}).   Special
cases related to NFW and MOA (Moore et al. 1999)
logarithmic density profiles, are plotted in
Fig.\,\ref{f:GPLs}.

\section{The RFSM5 method}
\label{res}

Given a set of SDH density profiles, one is left
with the problem of fitting a GPL density profile
to each simulation and to the average on the whole
set.   To this respect, nonlinear fits shall be 
used, minimizing the sum of both absolute values
and squares of absolute logarithmic residuals,
$R_i=\log\psi_{SDH}(\eta_i)-\log\psi_{GPL}(\eta_i)$,
used in the literature (e.g., Klypin et al. 2001;
Bullock et al. 2001).

Strictly speaking, the problem reduces to a search 
of extremum points of minimum, with regard to a
function, $Y=F(X_1,X_2,X_3,X_4,X_5)$, which
represents a 5-dimension hypersurface in a
6-dimension hyperspace.   In general, no point of
minimum can exist within the domain, if bounded,
or a finite number, or infinite.   If two or more
minima are present, then degeneracy occurs.   Owing to
an intrinsic difficulty related to the above
mentioned analytical procedure, a numerical
alternative shall be exploited.

More specifically, a 5-dimension hypergrid made
of $10^5$ points is placed around a fiducial
minimum, and the sum of both absolute values
and squares of absolute logarithmic residuals,
$\sum\vert R_i\vert$ and $\sum R_i^2$, respectively,
are evaluated at each point, and finally two
(in general) distinct minima are localized.   
Then a new iteration is performed, with respect
to a new hypergrid, centered near the minima,
where the size has been reduced and
superpositions on nodes of earlier hypergrids
have been avoided.   For the
calculations made in the current paper, three
iterations have been revealed to be sufficient.

In dealing with the hypergrid, it would be
better to use parameters with an immediate
geometrical meaning, instead of their analytical
counterparts, $(r_0,\rho_0,\alpha,\beta,\gamma)$.
To this aim, the logarithmic GPL density profile,
expressed by Eq.\,(\ref{eq:lpsi}), has to be
studied in detail, which is made in Appendix A.
The geometrical parameters to be
used, $(x_C,y_C,b,b_\beta,b_\gamma)$, are the
coordinates of the intersection of the asymptotes,
and the vertical intercepts of the curve and the
asymptotes, respectively.

In summary, the procedure outlined above acts
along the following steps.
\begin{description}
\item[\rm{(i)}\hspace{3.5mm}] Select a set of
SDH density profiles, related to an assigned
computer code and a specified cosmological
model.
\item[\rm{(ii)}~~] Determine, for each SDH 
density profile, the GPL density profiles
which minimize the sum of absolute values
and/or squares of absolute logarithmic
residuals, using a 5-dimension hypergrid
in the 5-dimension hyperspace $({\sf O}x_C
y_Cb~b_\beta b_\gamma)$.
\item[\rm{(iii)}~] Perform the desired 
number of iterations around the minima,
using a hypergrid with same number of
points (but none in common), and reduced
in size, in respect of its earlier analogon.
\item[\rm{(iv)}\hspace{1.7mm}] Define a  
mean SDH density profile, over the whole 
set of simulations, and repeat the procedure 
used for a single SDH density profile.
\end{description}
The result consists in a one-to-one
correspondence between SDH density profiles,
including the related mean,
and GPL density profiles which minimize the sum 
of absolute values and/or squares of absolute
logarithmic residuals, in connection with the
hypergrid used.   In the following, the above
mentioned procedure shall be quoted as RFSM5
(Residual Functions Sum Minimization within a
5-dimension hyperspace) method.   The functions
used in the current paper are the absolute
value and the square of the absolute logarithmic 
residual.

\section{An application to SDHs on the scale of
cluster of galaxies}\label{ara}

Using a RFSM5 method, GPL density profiles
are fitted to a sample of 17 SDH density
profiles, on the scale of cluster of
galaxies within a flat $\Lambda$CDM 
cosmology (Rasia et al. 2004, hereafter quoted as
RTM).    The values of the cosmological
parameters used therein are: $\Omega_\Lambda
=0.7$; $\Omega_m=0.3$; $\Omega_b=0.03$;
$h=0.7$; $\sigma_8=0.9$; where the symbols
have their usual captions (e.g., Klypin et
al. 2001; Bullock et al. 2001) and, in
particular, the indices $m$ and $b$ denote
all matter (dark + baryonic) and baryonic,
respectively.   For a detailed discussion on
the computer code, initial conditions, the
resolution issues, and the way of finding
the halo centre, see RTM and further
references therein.

Simulations include both dark and baryonic
matter, but only the former is relevant to 
the aim of the current paper.   Accordingly,
the baryonic matter shall not be considered,
and all the parameters shall be intended as
related to the dark matter halo.

The definition of the virialized region
within each halo, via Eq.\,(\ref{eq:Mvir}),
needs the knowledge of the critical
overdensity at virialization,
$\Delta_{{\rm vir}}$.   With regard to total
(dark + baryonic) matter, it depends on the
cosmological model (e.g., Bullock et al. 2001)
and, in the case under discussion, $(\Delta_
{vir})_m=323$ at $z=0$, where
all the sample haloes may be considered as
virialized to an acceptable extent (RTM).

If only the dark matter is considered, then
$(\Delta_{{\rm vir}})_d=\Delta_{{\rm vir}}=\zeta(\Delta_
{vir})_m$, where $\zeta$ is the fraction of dark
matter in each density perturbation, and
averaging over the whole sample yields $\zeta=
0.907$ (RTM).   Accordingly, the value:
\begin{equation}
\label{eq:Dvir}
\Delta_{{\rm vir}}=0.907\cdot323=292.961~~;
\end{equation}
can be used to an acceptable extent
\footnote{
The above value of the critical overdensity
was deduced from an earlier, unpublished version
of RTM.   It is slightly different from 
$\Delta_{{\rm vir}}=0.903\cdot323.7625=292.3576$,
deduced from the current version, which appeared 
when the calculations were performed in this paper.
As the relative difference amounts to about 0.2\%,
the calculations were not repeated using the
latter value of the critical overdensity.}.

As clearly pointed out in RTM, owing to the random
criterion used for selection, their
sample haloes are characterized by varying
dynamical properties: at the present time, some
are more relaxed, while others are dynamically
perturbed.   The surrounding environment can also
be quite different: some selected clusters are more
isolated, while others are interacting with the
surrounding cosmic web.   Accordingly, the
sample is expected to be good enough to provide
unbiased conclusions, and the related modelling
may be thought of as representative of an
average cluster, in an average
environment and dynamical configuration. 

\subsection{Individual SDH density profiles}
\label{iSDH}

The main features of sample haloes at
$z=0$ are listed in Tab.\,\ref{t:alvi}.
\begin{table}
\begin{tabular}{cccrrrr}
\hline
\hline
\multicolumn{1}{c}{case} &
\multicolumn{1}{c}{run} &
\multicolumn{1}{c}{type}  &
\multicolumn{1}{r}{$\phantom{28}N\phantom{74}$}  &
\multicolumn{1}{r}{$r_{{\rm vir}}
$} &
\multicolumn{1}{r}{$M_{{\rm vir}}
$} &
\multicolumn{1}{r}{$M_{{\rm vir}}^\prime
$} \\
\hline
\phantom{1}1 & $S01.02$ & $R$ & 282574 & 1953 &  76330 &  76040 \\
\phantom{1}2 & $S02.10$ & $M$ & 278569 & 2305 & 125400 & 125010 \\
\phantom{1}3 & $S02.11$ & $M$ &  85159 & 1553 &  38340 &  38234 \\
\phantom{1}4 & $S03.05$ & $R$ & 294373 & 2347 & 132500 & 131970 \\
\phantom{1}5 & $S04.01$ & $R$ & 179681 & 1991 &  80820 &  80565 \\
\phantom{1}6 & $S04.07$ & $R$ & 146386 & 1860 &  65850 &  65686 \\
\phantom{1}7 & $S05.02$ & $M$ & 318653 & 2197 & 108700 & 108249 \\
\phantom{1}8 & $S06.01$ & $M$ & 427583 & 2470 & 153800 & 153825 \\
\phantom{1}9 & $S06.03$ & $M$ & 166855 & 1796 &  60020 &  59136 \\
10           & $S07.01$ & $R$ & 275259 & 1691 &  49600 &  49359 \\
11           & $S07.03$ & $R$ & 158345 & 1407 &  28530 &  28433 \\
12           & $S08.01$ & $R$ & 190453 & 1884 &  68600 &  68262 \\
13           & $S08.04$ & $R$ & 101482 & 1529 &  36560 &  36489 \\
14           & $S09.03$ & $R$ & 159330 & 1913 &  71690 &  71463 \\
15           & $S09.14$ & $R$ & 107229 & 1675 &  48250 &  47971 \\
16           & $S10.03$ & $R$ &  58734 & 1524 &  36060 &  36132 \\
17           & $S10.07$ & $R$ &  71937 & 1628 &  44170 &  44045 \\
\hline
\end{tabular}
\caption{Main features of sample haloes at
$z=0$.   Column captions: 1 - case; 2 - 
computer run; 3 - type ($R$ - safely relaxed; 
$M$ - safely
a major merger occurring); 4 - number of dark
matter particles within the virial radius;
5 - virial radius ($h^{-1}~$kpc); 6 - virial
mass ($h^{-1}10^{10}{\rm M}_\odot$); 7 - virial
mass deduced from Eq.\,(\ref{eq:Mvir}).   Both 
virial radii and virial masses are normalized
to the dimensionless Hubble parameter at the
current time, $h$.}
\label{t:alvi}
\end{table}
The mass, $M_{{\rm vir}}$, has been taken from
RTM, while the mass, $M_{{\rm vir}}^\prime$,
has been deduced from Eq.\,(\ref{eq:Mvir}).
The apparent discrepancy between $M_{{\rm vir}}$
and $M_{{\rm vir}}^\prime$ is owing to two
different sources.   First, a systematic
contribution takes origin from the
uncertainty on $\Delta_{{\rm vir}}$ and, second,
a random contribution arises from the
uncertainty on $r_{{\rm vir}}$, in both cases with
regard to Eq.\,(\ref{eq:Mvir}).   An
additional random contribution is related
to averaging the fraction of dark matter
over the whole sample, with regard to
Eq.\,(\ref{eq:Dvir}).

The relative difference, $\vert1-M_{{\rm vir}}^
\prime/M_{{\rm vir}}\vert$, is less than one
percent in all cases except 9, where it 
is less than one and half percent.   Then
the virial mass can be evaluated, to a good 
extent, by use of Eq.\,(\ref{eq:Mvir}),
taking the virial radius from the results of 
simulations.
It is worth mentioning that the RFSM5 method
is independent of the value of the virial
mass, while a change in the value of the
virial radius makes SDH density profiles
systematically shift along the horizontal 
direction, see Fig.\,\ref{f:SDH}.

\subsection{Averaged SDH density profiles}
\label{aSDH}               

Given a set of logarithmic SDH density 
profiles, the mean SDH density profile
is obtained by averaging over the whole
set the values related to each logarithmic
radial bin, in the range of interest, 
expressed by Eq.\,(\ref{eq:doeta}).     
The value of the critical overdensity
at virialization, $\Delta_{{\rm vir}}$,
is fixed by Eq.\,(\ref{eq:Dvir}), then a
single free parameter remains: the virial
radius, $r_{{\rm vir}}$, which allows the
calculation of the virial mass, $M_{{\rm vir}}$.
The related values are determined by
averaging over the whole sample, using
the data listed in Tab.\,\ref{t:alvi},
and inserting the mean value of the virial 
radius into Eq.\,(\ref{eq:Mvir}).   The
result is:
\begin{equation}
\label{eq:rMvm}
\overline{r}_{{\rm vir}}=1866~h^{-1}~{\rm kpc}~~;
\qquad\overline{M}_{{\rm vir}}=66330~h^{-1}~10^{10}
{\rm M}_\odot~~;
\end{equation}
and the application of a RFSM5 method
yields best fitting GPL density profiles,
with radius and mass equal to $\overline{r}_
{vir}$ and $\overline{M}_{{\rm vir}}$, respectively.

A mean virial radius has been preferred
in place of a mean concentration (RTM)
for the following reasons.   First,
virial radii are independent of GPL
density profiles, contrary to concentrations,
or velocity profiles, which should be
calculated for any choice of the fitting
profile.   Second, the range of virial
radii, $1407\le r_{{\rm vir}}/(h^{-1}{\rm kpc})\le
2470$, corresponds to relative errors of about 
25\% and 33\%, respectively, with regard
to a mean value, $\overline{r}_{{\rm vir}}=1866~h^
{-1}$kpc.   On the other hand, the range of
concentrations (calculated for NFW density
profiles), $5\le\xi_{{\rm vir}}\le10$,
corresponds to relative errors of about
32\% and 37\%, respectively, with regard
to a mean value, $\overline{\xi}_{{\rm vir}}=7.2976$.
Then the average of the virial radius
should be preferred to this respect.

Having in our hands a SDH density profile
averaged over the sample, listed in 
Tab.\,\ref{t:alvi}, and a value of virial
radius and virial mass, expressed by
Eq.\,(\ref{eq:rMvm}), we are left with
the search of a best fitting GPL density
profile.  To this aim, six alternatives
are exploited.   The first one consists
in the mere application of the RFSM5
method to the mean SDH density profile.

Among the remaining five, three allow
to fix the exponents in the GPL density
profile, expressed by Eq.\,(\ref{eq:GPL}),
and then minimize the sum of absolute
values and/or squares of absolute
logarithmic residuals, with respect to
a single free parameter, the scaling
radius, $r_0$ (e.g., Zhao et al. 2003).
Accordingly, the minimization is
performed using a 2-dimension grid.
The related procedure shall be quoted
as RFSM2 (Residual Functions Sum  
Minimization within a 2-dimension space)
method.   The function used in the
current paper is the absolute value
and/or the square of the absolute
logarithmic residual.   The following
special GPL density profiles are
selected: NFW, MOA, and a best
fitting profile deduced from both
density and velocity distributions
in sample haloes (RTM), hereafter
quoted as RTM density profile
\footnote{It is worth mentioning
that a different normalization has
been used here for the scaling
density, $\rho_0=(\xi_{{\rm vir}})^{5/2}
(\rho_0)_{RTM}$.}.%
The related values of the exponents 
are $(\alpha,\beta,\gamma)=$(1, 3, 1),
(3/2, 3, 3/2), (1, 5/2, 1), respectively.

Finally, the last two alternatives
among the six mentioned above,
consist in calculating the mean
values of the geometrical parameters,
$(x_C,y_C,b,~b_\beta,b_\gamma)$, over
the whole sample of best fitting, GPL
density profiles, with regard to the
minimization of the sum of absolute
values and squares of absolute 
logarithmic residuals, respectively.

\subsection{Results}
\label{resu}

As outlined in section \ref{res}, 
a RFSM5 method has been applied to each 
sample halo, listed in Tab.\,\ref{t:alvi},
and to the related, averaged SDH density
profile, which has been defined above.
In addition, a RFSM2 method has been
applied to the mean SDH density profile,
in the special case of NFW, MOA, and
RTM density profiles.   The values of
the exponents, $\alpha$, $\chi$, $\beta$,
$\gamma$, the scaled radius, $\xi_{{\rm vir}}$,
the scaling radius, $r_0$, the scaling
density, $\rho_0$, and the sum of residual
functions
at the fiducial minimum, $\sum f(R_i)$,
where $f(R_i)=\vert R_i\vert$ and $f(R_i)
=R_i^2$, are listed in Tabs.\,\ref{t:para} 
and \ref{t:parq}, respectively.
The following conclusions are deduced.
\begin{table}
\begin{tabular}{cllllrlll}
\hline
\hline
\multicolumn{1}{c}{case} &
\multicolumn{1}{c}{$\alpha$} & 
\multicolumn{1}{c}{$\chi$} & 
\multicolumn{1}{c}{$\beta$} &
\multicolumn{1}{c}{$\gamma$} &
\multicolumn{1}{c}{$\xi_{{\rm vir}}$} &
\multicolumn{1}{c}{$r_0$} &
\multicolumn{1}{c}{$10^4\rho_0$} &
\multicolumn{1}{c}{$\sum\vert R_i\vert$} \\
\hline
\phantom{1}1 & 0.74154 & 3.1652 & 3.4698 & 1.1226 & 3.3884 & 823.39 &
0.51314 & 0.37127 \\
\phantom{1}2 & 0.55959 & 4.5587 & 3.3698 & 0.81879 & 3.9446 & 834.78 &
1.6992 & 0.67590 \\
\phantom{1}3 & 0.44184 & 3.7405 & 2.8307 & 1.1780 & 6.6069 & 335.79 & 
2.6317 & 0.77869 \\
\phantom{1}4 & 0.45956 & 3.6948 & 2.7388 & 1.0409 & 7.2577 & 461.97 &
2.6930 & 0.76086 \\
\phantom{1}5 & 0.65993 & 2.7518 & 3.0723 & 1.2563 & 4.7315 & 601.14 &
0.74171 & 0.45363 \\
\phantom{1}6 & 0.52739 & 3.2392 & 2.6766 & 0.96825 & 13.884\phantom{1} 
& 191.38 & 7.3923 & 0.62044 \\
\phantom{1}7 & 0.59587 & 4.0247 & 3.4760 & 1.0778 & 4.6559 & 674.11 &
1.8207 & 0.57012 \\
\phantom{1}8 & 0.66861 & 3.9138 & 3.4748 & 0.85806 & 4.1687 & 846.45 &
1.3190 & 0.22446 \\
\phantom{1}9 & 0.56370 & 3.9028 & 3.1506 & 0.95062 & 6.4565 & 397.38 &
3.2377 & 0.27807 \\
10 & 0.42473 & 3.8973 & 2.7797 & 1.1139 & 6.1660 & 391.78 &
2.3455 & 0.47658 \\
11 & 0.65178 & 3.7207 & 3.5506 & 1.1255 & 3.2359 & 621.15 & 
0.67645 & 0.44895 \\
12 & 0.66423 & 3.5417 & 2.9664 & 0.61389 & 7.9433 & 338.83 &
3.2752 & 0.75673 \\
13 & 0.70250 & 3.1223 & 3.1281 & 0.93462 & 6.0256 & 362.50 &
1.6605 & 0.58189 \\
14 & 0.74575 & 3.2092 & 3.2130 & 0.81972 & 5.1286 & 532.86 & 
1.2029 & 0.35644 \\
15 & 0.82346 & 2.1323 & 2.7865 & 1.0307 & 6.5313 & 366.37 &
0.91250 & 0.59466 \\
16 & 0.71709 & 3.8681 & 3.7082 & 0.93443 & 4.0738 & 534.43 &
1.2889 & 0.43596 \\
17 & 0.47209 & 3.1773 & 2.6513 & 1.1513 & 5.7544 & 404.16 & 
1.1488 & 0.66217 \\
\hline
ADP & 0.54955 & 4.5235 & 3.3962 & 0.91034 & 3.8019 & 701.18 &
1.5496 & 0.12966 \\
AGP & 0.60209 & 3.4021 & 3.0756 & 1.0272 & 5.5083 & 483.96 &
1.5422 & 0.36819 \\
NFW & 1 & 2 & 3 & 1 & 6.35 & 419.81 & 0.90131 & 0.85996 \\
MOA & 1.5 & 1 & 3 & 1.5 & 3.08 & 865.52 & 0.093933 & 0.47639 \\
RTM & 1 & 1.5 & 2.5 & 1 & 13.050\phantom{1} & 204.28 & 
2.1949 & 0.32672 \\
\hline\hline
\end{tabular}
\caption{Parameters of GPL density profiles which
(i) minimize the sum of absolute values of logarithmic 
absolute residuals, using a RFSM5 method with
regard to 17 sample haloes listed in Tab.\,\ref{t:alvi}
(top), and (ii) fit to the mean SDH density profile,
to a different extent (bottom).   Cases correspond to
computer runs in the former alternative, and to GPL
density profiles in the latter.   The GPL density
profile which minimizes the sum of absolute values of
absolute logarithmic
residuals, using a RFSM5 method with regard to the mean
SDH density profile, is denoted as ADP.   The GPL
density profile defined by geometrical parameters, 
$(x_C,y_C,b,b_\beta,b_\gamma)$, averaged over their
counterparts listed on the top (first 17 rows), is
denoted as AGP.   The scaling radius, $r_0$, and the
scaling density, $\rho_0$, are expressed in kpc and
$10^{10}{\rm M}_\odot/{\rm kpc}^3$, respectively.}
\label{t:para}
\end{table}
\begin{table}
\begin{tabular}{cllllrlll}
\hline
\hline
\multicolumn{1}{c}{case} &
\multicolumn{1}{c}{$\alpha$} & 
\multicolumn{1}{c}{$\chi$} & 
\multicolumn{1}{c}{$\beta$} &
\multicolumn{1}{c}{$\gamma$} &
\multicolumn{1}{c}{$\xi_{{\rm vir}}$} &
\multicolumn{1}{c}{$r_0$} &
\multicolumn{1}{c}{$10^4\rho_0$} &
\multicolumn{1}{c}{$\sum R_i^2$} \\
\hline
\phantom{1}1 & 0.85979 & 2.7291 & 3.4893 & 1.1429 & 3.6308 & 768.43 &
0.45733 & 0.012843 \\
\phantom{1}2 & 0.70267 & 2.6771 & 2.8238 & 0.94262 & 7.1450 & 460.86 &
1.4462 & 0.050984 \\
\phantom{1}3 & 0.50594 & 3.6824 & 3.0631 & 1.2000 & 4.4668 & 496.68 & 
1.1755 & 0.059798 \\
\phantom{1}4 & 0.40157 & 4.8960 & 2.9564 & 0.99023 & 4.1115 & 815.48 &
2.1391 & 0.044834 \\
\phantom{1}5 & 0.59295 & 3.2455 & 3.1244 & 1.2000 & 4.7863 & 594.26 &
1.0721 & 0.015935 \\
\phantom{1}6 & 0.50437 & 2.8516 & 2.5152 & 1.0769 & 16.501\phantom{1} & 
161.03 & 7.3632 & 0.033055 \\
\phantom{1}7 & 0.60451 & 4.1915 & 3.5715 & 1.0377 & 4.2267 & 742.56 &
1.7070 & 0.029137 \\
\phantom{1}8 & 0.66861 & 3.9138 & 3.4748 & 0.85806 & 4.1687 & 846.45 &
1.3190 & 0.0049367 \\
\phantom{1}9 & 0.58477 & 3.7717 & 3.2083 & 1.0027 & 5.6234 & 456.26 &
2.2399 & 0.0061089 \\
10 & 0.43019 & 3.7611 & 2.7372 & 1.1192 & 6.6374 & 363.95 &
2.5132 & 0.020731 \\
11 & 0.69931 & 3.3055 & 3.4625 & 1.1509 & 3.7154 & 541.00 & 
0.67645 & 0.017674 \\
12 & 0.67451 & 3.0786 & 2.7795 & 0.70312 & 9.1201 & 363.19 &
2.8856 & 0.050487 \\
13 & 0.68562 & 3.3840 & 3.1595 & 0.83933 & 6.8234 & 320.12 &
2.5718 & 0.036570 \\
14 & 0.81937 & 2.4907 & 2.9870 & 0.94621 & 5.3088 & 514.77 & 
0.77666 & 0.012273 \\
15 & 0.70472 & 2.8497 & 2.8694 & 0.86117 & 7.6913 & 311.11 &
1.9963 & 0.028821 \\
16 & 0.74414 & 3.9381 & 3.8674 & 0.93684 & 3.7154 & 585.99 &
1.1226 & 0.015259 \\
17 & 0.44450 & 4.0781 & 2.8351 & 1.0224 & 4.6774 & 497.23 & 
1.5144 & 0.057718 \\
\hline
ADP & 0.56832 & 4.0722 & 3.3143 & 1.0000 & 3.6308 & 734.22 &
1.0238 & 0.0014983 \\
AGP & 0.58866 & 3.5008 & 3.0528 & 0.99204 & 5.5170 & 483.20 &
1.5726 & 0.0065805 \\
NFW & 1 & 2 & 3 & 1 & 6.44 & 413.94 & 0.93151 & 0.046088 \\
MOA & 1.5 & 1 & 3 & 1.5 & 3.05 & 874.03 & 0.091827 & 0.015492 \\
RTM & 1 & 1.5 & 2.5 & 1 & 13.350\phantom{1} & 199.69 & 
2.3074 & 0.0072902 \\
\hline\hline
\end{tabular}
\caption{Parameters of GPL density profiles which
(i) minimize the sum of squares of logarithmic 
absolute residuals, using a RFSM5 method with
regard to 17 sample haloes listed in Tab.\,\ref{t:alvi}
(top), and (ii) fit to the mean SDH density profile,
to a different extent (bottom).   Cases correspond to
computer runs in the former alternative, and to GPL
density profiles in the latter.   The GPL density
profile which minimizes the sum of squares of
absolute logarithmic
residuals, using a RFSM5 method with regard to the mean
SDH density profile, is denoted as ADP.   The GPL
density profile defined by geometrical parameters, 
$(x_C,y_C,b,b_\beta,b_\gamma)$, averaged over their
counterparts listed on the top (first 17 rows), is
denoted as AGP.   The scaling radius, $r_0$, and the
scaling density, $\rho_0$, are expressed in kpc and
$10^{10}{\rm M}_\odot/{\rm kpc}^3$, respectively.}
\label{t:parq}
\end{table}
\begin{description}
\item[\rm{(i)}\hspace{3.5mm}] In general,
different GPL density profiles best fit to
an assigned SDH density profile, with regard
to the minimization of the sum of absolute
values or squares of absolute logarithmic
residuals, respectively.   The best fitting
GPL density profiles in the two above mentioned
alternatives, are closer each to the other in
dealing with the mean geometrical parameters,
than with the mean SDH density profile.
\item[\rm{(ii)}~~] The values of the exponents,
$(\alpha,\beta,\gamma)$, appearing in Eq.\,(\ref
{eq:GPL}), and deduced from the geometrical
parameters averaged over the whole halo sample, 
$(\overline{x}_C,\overline{y}_C,\overline{b},
\overline{b}_\beta,\overline{b}_\gamma)$, are 
$(\overline{\alpha},\overline{\beta},
\overline{\gamma})\approx(0.6,3.1,1.0)$.
\item[\rm{(iii)}~] With regard to the mean SDH
density profile, the GPL density profiles which
best minimize the sum of squares of absolute
logarithmic residuals, occur in the following
order of accuracy: (1) application of a RFSM5 
method (ADP); (2) density profile related to 
mean values of geometrical parameters (AGP);
(3) RTM; (4) MOA; (5) NFW.   The best
minimization of the sum of absolute values of 
absolute logarithmic residuals, implies the
same order as above, but with AGP and RTM
density profiles interchanged.
\end{description}

\begin{figure}
\centerline{\psfig{file=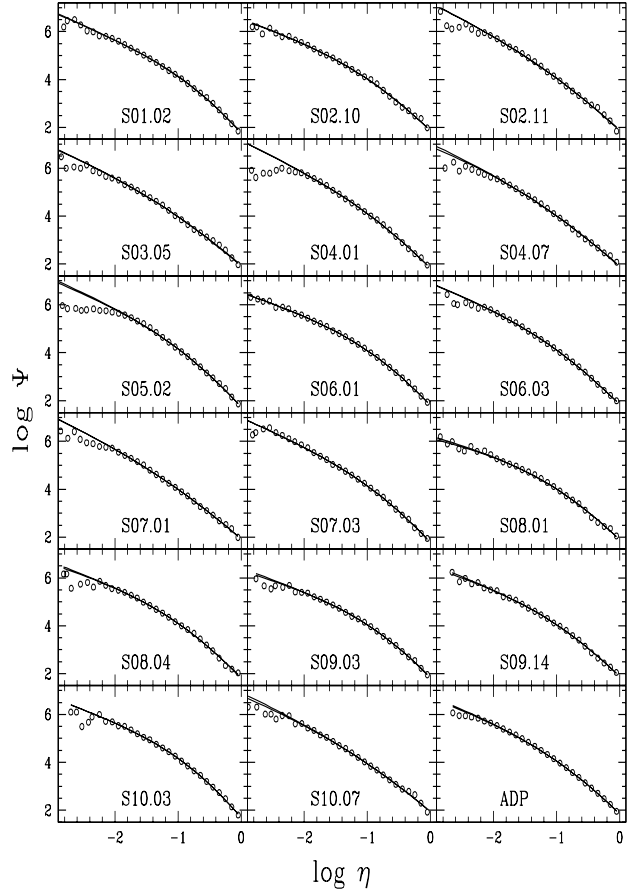,height=130mm,width=90mm}}
\caption{The SDH density profiles related to the 
current sample, listed in Tab.\,\ref{t:alvi}, and 
the mean SDH density profile, denoted as $ADP$
(open circles), together with their best fitting GPL 
counterparts (full curves), determined by use of a
RFSM5 method in the range $-2<\log(r/r_{{\rm vir}})<0$.
Two curves on each panel correspond to the minimization
of the sum of absolute values and squares of
absolute logarithmic residuals, respectively.   The
above mentioned curves are virtually indistinguishable
in most cases, and sometimes coincident.}
\label{f:pd18} 
\end{figure}

The SDH density profiles related to the
current sample, listed in Tab.\,\ref{t:alvi},
and the related mean SDH density profile,
are plotted in Fig.\,\ref{f:pd18} (dots)
together with their best fitting GPL counterpart
(full curves),
determined by use of a RFSM5 method in the 
range defined by Eq.\,(\ref{eq:doeta}).
In most cases, GPL density profiles related
to the minimization of the sum of absolute
values and squares of absolute logarithmic
residuals, are virtually indistinguishable,
and sometimes coincident.

\begin{figure}
\centerline{\psfig{file=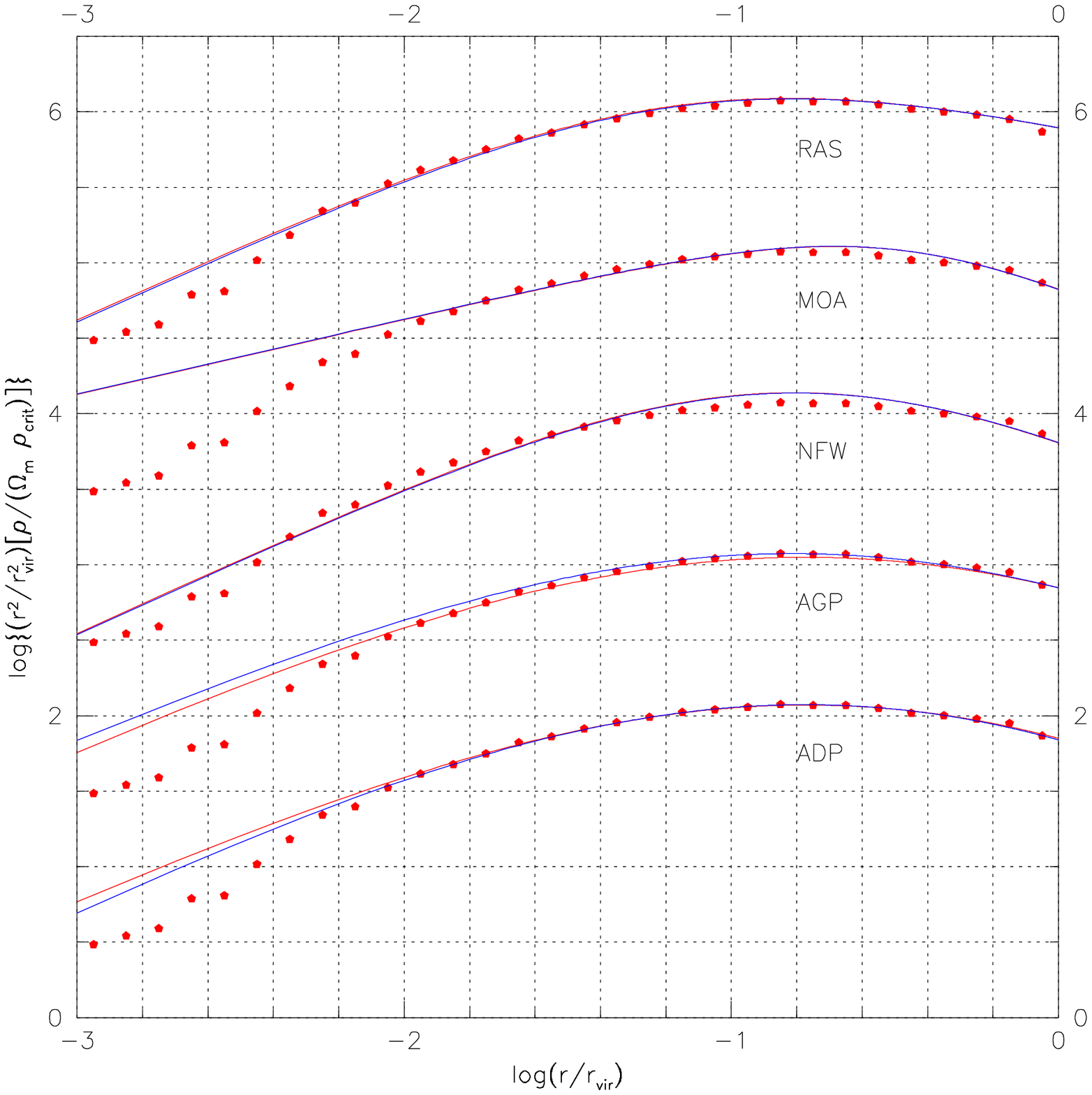,height=130mm,width=90mm}}
\caption{Comparison between different fits 
(full curves) to the mean SDH density profiles 
(filled circles), listed on the lower parts of
Tabs.\,\ref{t:para} and \ref{t:parq}.
The function $\log[(r/r_{{\rm vir}})^2(\rho/\rho_h)]$
has been plotted instead of $\log(\rho/\rho_h)$, 
to make different trends more evident.   The
vertical scale is related to the lower case.
The remaining cases are, in turn, vertically
shifted of one unity with respect to their
closest counterpart, starting from the
lower one, to gain clarity.   
With regard to case AGP,  the upper and the lower
curve correspond to the minimization
of the sum of absolute values and squares of
absolute logarithmic residuals, respectively,
and the contrary holds for case ADP.   Concerning
the remaining cases NFW, MOA, and RTM (labelled
here as RAS), the two curves are virtually
indistinguishable.}
\label{f:pdm5} 
\end{figure}

Different fits to the mean SDH density
profile, listed on the lower parts of
Tabs.\,\ref{t:para} and \ref{t:parq}, 
are represented in Fig.\,\ref{f:pdm5},
where $\log[(r/r_{{\rm vir}})^2(\rho/\rho_h)]$
has been plotted instead of $\log(\rho/
\rho_h)$, to make different trends more
evident.
Curves related to the minimization of
the sum of absolute values and squares
of absolute logarithmic residuals are
virtually indistinguishable with regard
to GPL density profiles with fixed 
exponents: NFW, MOA, and RTM.

The values of some analytical and geometrical
parameters, $\eta_{ADP}$, related to the
best fitting GPL density profile to the mean
SDH density profile, are listed in Tabs.\,\ref
{t:parma} and \ref{t:parmq} together with 
their counterparts, $\overline{\eta}$, averaged
over the best fitting GPL density profiles
to the whole halo sample, via minimization
of the sum of absolute values and squares
of absolute logarithmic residuals, respectively.   
\begin{table}
\begin{tabular}{llllll}
\hline
\hline
\multicolumn{1}{c}{$\eta$} &
\multicolumn{1}{c}{$\eta_{ADP}$} & 
\multicolumn{1}{c}{$\overline{\eta}$} & 
\multicolumn{1}{c}{$\sigma_{s~\eta}$} &
\multicolumn{1}{c}{$\sigma_{s~\overline{\eta}}$} &
\multicolumn{1}{c}{$\sigma_{s~\overline{\mu}}$} \\

\hline
$\alpha$        & $\phantom{-}$0.54955  & 0.61292  & 0.11931  & 0.028937 & 0.0049626 \\
$\beta$         & $\phantom{-}$3.3962   & 3.1202   & 0.34247  & 0.083062 & 0.014245  \\
$\gamma$        & $\phantom{-}$0.91034  & 0.99973  & 0.16188  & 0.039261 & 0.0067332 \\
$\xi_{{\rm vir}}$     & $\phantom{-}$3.8019   & 5.8796   & 2.4766   & 0.60068  & 0.10301   \\
$\nu_{mas}$     & $\phantom{-}$0.42346  & 1.4063   & 1.0804   & 0.26205  & 0.044941  \\
$\log\xi_{{\rm vir}}$ & $\phantom{-}$0.58     & 0.74102  & 0.15471  & 0.037522 & 0.0064350 \\
$\log\nu_{mas}$ & $\phantom{ }-$0.37319 & 0.050834 & 0.29116  & 0.070617 & 0.012111  \\
$y_C$           & $\phantom{-}$4.58     & 4.5779   & 0.29347  & 0.071178 & 0.012207  \\
$b$             & $\phantom{-}$1.84     & 1.8467   & 0.065779 & 0.015954 & 0.0027360 \\
$b_\beta$       & $\phantom{-}$2.6102   & 2.2988   & 0.14859  & 0.036038 & 0.0061804 \\
$b_\gamma$      & $\phantom{-}$4.052    & 3.8167   & 0.31970  & 0.077538 & 0.013298  \\
\hline\hline
\end{tabular}
\caption{Comparison between parameters, $\eta
_{ADP}$, related to the best fitting GPL
density profile to the mean SDH density 
profile, and their counterparts, $\overline{\eta}$,
averaged over the best fitting GPL density
profiles to the whole halo sample, with regard 
to the minimization of the sum of absolute
values of absolute logarithmic residuals.
Also listed are the related standard deviations,
$\sigma_{s~\eta}$, the standard deviations from
the mean, $\sigma_{s~\overline{\eta}}$, and the
standard deviations from the standard deviation
from the mean, $\sigma_{s~\overline{\mu}}$.   It is
worth remembering that $\log\xi_{{\rm vir}}=-x_C$,
according to Eq.\,(\ref{eq:xC}).}
\label{t:parma}
\end{table}
\begin{table}
\begin{tabular}{llllll}
\hline
\hline
\multicolumn{1}{c}{$\eta$} &
\multicolumn{1}{c}{$\eta_{ADP}$} & 
\multicolumn{1}{c}{$\overline{\eta}$} & 
\multicolumn{1}{c}{$\sigma_{s~\eta}$} &
\multicolumn{1}{c}{$\sigma_{s~\overline{\eta}}$} &
\multicolumn{1}{c}{$\sigma_{s~\overline{\mu}}$} \\

\hline
$\alpha$        & $\phantom{-}$0.56832  & 0.62515 & 0.13353  & 0.032860 & 0.0055541 \\
$\beta$         & $\phantom{-}$3.3143   & 3.1132  & 0.35867  & 0.086990 & 0.014919 \\
$\gamma$        & $\phantom{-}$1.0000   & 1.0018  & 0.13893  & 0.033696 & 0.0057788 \\
$\xi_{{\rm vir}}$     & $\phantom{-}$3.6308   & 6.0210  & 3.1377   & 0.76101  & 0.13051   \\
$\nu_{mas}$     & $\phantom{-}$0.55823  & 1.7340  & 1.6790   & 0.40721  & 0.069837  \\
$\log\xi_{{\rm vir}}$ & $\phantom{-}$0.56     & 0.74172 & 0.17256  & 0.041852 & 0.0071775 \\
$\log\nu_{mas}$ & $\phantom{ }-$0.25319 & 0.10544 & 0.33340  & 0.080860 & 0.013867  \\
$y_C$           & $\phantom{-}$4.4      & 4.5864  & 0.27926  & 0.067732 & 0.011616  \\
$b$             & $\phantom{-}$1.85     & 1.8480  & 0.075648 & 0.018347 & 0.0031465 \\
$b_\beta$       & $\phantom{-}$2.544    & 2.3221  & 0.21914  & 0.053149 & 0.0091151 \\
$b_\gamma$      & $\phantom{-}$3.84     & 3.8506  & 0.21627  & 0.052454 & 0.0089958 \\
\hline\hline
\end{tabular}
\caption{Comparison between parameters, $\eta
_{ADP}$, related to the best fitting GPL
density profile to the mean SDH density 
profile, and their counterparts, $\overline{\eta}$,
averaged over the best fitting GPL density
profiles to the whole halo sample, with regard 
to the minimization of the sum of squares
of absolute logarithmic residuals.
Also listed are the related standard deviations,
$\sigma_{s~\eta}$, the standard deviations from
the mean, $\sigma_{s~\overline{\eta}}$, and the
standard deviations from the standard deviation
from the mean, $\sigma_{s~\overline{\mu}}$.   It is
worth remembering that $\log\xi_{{\rm vir}}=-x_C$,
according to Eq.\,(\ref{eq:xC}).}
\label{t:parmq}
\end{table}
Also listed therein are
the related standard deviations, $\sigma_
{s~\eta}$, the standard deviations from
the mean, $\sigma_{s~\overline{\eta}}$, and the
standard deviations from the standard
deviation from the mean, $\sigma_{s~\overline
{\mu}}$, which are expressed as (e.g., Oliva \&
Terrasi 1976, Chap.\,V, \S\,5.6.3):
\begin{lefteqnarray}
\label{eq:etam}
&& \overline{\eta}=\frac1n\sum_{i=1}^n\eta_i~~; \\
\label{eq:seta}
&& \sigma_{s{\eta}}=\left[\frac1{n-1}\sum_
{i=1}^n(\eta_i-\overline{\eta})^2\right]^{1/2}~~; \\
\label{eq:setam}
&& \sigma_{s\overline{\eta}}=\left[\frac1n\frac1{n-1}\sum_
{i=1}^n(\eta_i-\overline{\eta})^2\right]^{1/2}~~; \\
\label{eq:smum}
&& \sigma_{s\overline{\mu}}=\frac{\sigma_{s\overline{\eta}}}
{\sqrt{2n}}~~;\quad\overline{\mu}=\sigma_{s\overline{\eta}}~~;
\end{lefteqnarray}
where $n=17$; $\eta=\alpha$, $\beta$, $\gamma$,
$\xi_{{\rm vir}}$, $\nu_{mas}$, $\log\xi_{{\rm vir}}$, 
$\log\nu_{mas}$, $y_C$, $b$, $b_\beta$, $b_\gamma$;
and, owing to Eq.\,(\ref{eq:xC}), $\log\xi_{{\rm vir}}=-x_C$.

The following conclusions are deduced from 
Tabs.\,\ref{t:parma} and \ref{t:parmq}.
\begin{description}
\item[\rm{(iv)}\hspace{1.7mm}] Values of 
parameters, $\eta_{ADP}$, related to
the best fitting GPL density profile to
the mean SDH density profile, are different
from their counterparts averaged over the
best fitting GPL density profiles to the
whole halo sample, as expected from the 
theory of errors.
\item[\rm{(v)}~] The exponents of best
fitting, GPL density profiles, are close
to their NFW counterparts, conform to
$[\Nint(\alpha),\Nint(\beta),\Nint(\gamma)]
=(1,3,1)$, where $\Nint$ denotes the
nearest integer.   The difference increases
from about one hundredth for $\gamma$
to about one tenth for $\beta$, and to
about one half for $\alpha$.
\item[\rm{(vi)}\hspace{1.0mm}] The rms
error of the logarithm of the scaled radius,
$\xi_{{\rm vir}}$, is $\sigma_{s~\log\xi_{{\rm vir}}}=
0.15-0.17$, to be compared with $\sigma_{s~
\log\xi_{{\rm vir}}}=0.18$ deduced from richer
samples where $(\alpha,\beta,\gamma)=(1,3,1)$;
$M_{{\rm vir}}=$(0.5-1.0)$\times10^nh^{-1}{\rm m}_
\odot$; $11\le n\le14$; and $n$ is an integer
(Bullock et al. 2001).
\end{description}

\subsection{Discussion}\label{disc}

The application of a RFSM5 method suceeds
in minimizing the sum of absolute values
and squares of absolute logarithmic residuals,
with respect to GPL density profiles where
the exponents are kept fixed, such as NFW,
MOA, and RTM, which allows the following 
definition.
\begin{trivlist}
\item[\hspace\labelsep{\bf Universal density
profile.}] \sl
Let a RFSM5 method be applied to an assigned
set of SDH density profiles.   The best
fitting GPL density profile to the mean
SDH density profile, is defined as the
related universal density profile.
\end{trivlist}
In this view, ``universal'' has to be
intended as nothing but ``best fitting''.
Strictly speaking, the above statement
should apply to a simulated halo sample
which is representative of the whole set
of real dark matter haloes.

The minimization of the sum of absolute
values or squares of absolute logarithmic
residuals, makes a firm criterion for   
deciding which, among two or more GPL
density profiles, best fits an assigned
SDH density profile.   The results of
the current paper confirm earlier results
about sample haloes on the scale of
cluster of galaxies, namely (a) MOA
density profiles provide a better fit
with respect to NFW (Fukushige \&
Makino 2001, 2003), and (b) RTM density
profiles provide a better fit with
respect to NFW (RTM), with the
additional result (c) RTM density
profiles provide a better fit with
respect to MOA.

A RFSM5 method has recently been used
for determining the scaled radius
(Hiotelis 2003), but
using a different definition with respect
to $\xi_{{\rm vir}}=r_{{\rm vir}}/r_0$, Eq.\,(\ref
{eq:csiv}).   In fact, the usual definition
of concentration is $c=r_{{\rm vir}}/r_{-2}$, 
where $r_{-2}$ is the radius related to a
logarithmic slope, defined by Eq.\,(\ref
{eq:dyx}), $\diff y/\diff x=-2$ (e.g.,
Klypin et al. 2001; Bullock et al. 2001;
Hiotelis 2003).   The former definition
seems to be more general, as it allows
the maximum change in slope at the scaled
radius, as pointed out in section \ref{GPL}.
In addition, it makes the definition of
concentration meaningful also in early
times, where the slope of a GPL density
profile may be smaller (in absolute value)
than 2 (Hiotelis 2003).

In dealing with dark matter haloes on the
scale of cluster of galaxies, Hiotelis
(2003) finds GPL density profiles where
the exponent, $\gamma$, attains a value
of about 1.5, in contrast with the results
of the current paper, $\gamma\approx1$.
Such a discrepancy is probably owing to
the different definitions of concentration,
mentioned above, which have been used.

On the other hand, values of asymptotic
inner slopes of fitting density profiles
determined in the current
paper, are consistent with their counterparts
deduced from recent high-resolution
simulations using a three-parameter fit
involving scaling radius, scaling density,
and asymptotic inner slope (6 sample objects,
Diemand et al. 2004) or a two-parameter fit
involving scaling radius and asymptotic inner
slope (16 sample objects, Reed et al. 2005).
The related parameters are listed in Tab.\,\ref
{t:gamma}, which shows agreement between
different approaches, within the fiducial range,
$\overline{\gamma}\mp3\sigma_{s~\overline{\gamma}}$.
\begin{table}
\begin{tabular}{clllllllll}
\hline
\hline
\multicolumn{1}{c}{c} &
\multicolumn{1}{c}{$n$} &
\multicolumn{1}{c}{$\overline{\gamma}$} &
\multicolumn{1}{c}{$\sigma_{s~\gamma}$} &
\multicolumn{1}{c}{$\sigma_{s~\overline{\gamma}}$} &
\multicolumn{1}{c}{$\Delta^-\gamma$}  &
\multicolumn{1}{c}{$\Delta^+\gamma$} &
\multicolumn{1}{c}{$\Delta^\mp\gamma$} &
\multicolumn{1}{c}{$\overline{\gamma}^-$} &
\multicolumn{1}{c}{$\overline{\gamma}^+$} \\

\hline
A &            17  & 0.99973 & 0.16188 & 0.039261 & 0.38584 & 0.25657 & 0.32120 & 0.88195 & 1.1175 \\
S &            17  & 1.0018  & 0.13893 & 0.033696 & 0.29868 & 0.19820 & 0.24844 & 0.90071 & 1.1029 \\
D & $\phantom{1}6$ & 1.1617  & 0.13732 & 0.056060 & 0.24167 & 0.25833 & 0.25000 & 0.99352 & 1.3299 \\
R &            16  & 1.2875  & 0.23910 & 0.059774 & 0.28750 & 0.41250 & 0.35000 & 1.1082  & 1.4668 \\
R &            13  & 1.2692  & 0.24285 & 0.067353 & 0.26923 & 0.43077 & 0.35000 & 1.0671  & 1.4713 \\
\hline\hline
\end{tabular}
\caption{Comparison between statistical parameters related
to the asymptotic inner slope, $\gamma$, deduced from
different samples using different fits.   Column captions:
1 - case: A - current paper, minimization of the sum of
absolute values of absolute logarithmic residuals; S -
current paper, minimization of the sum of squares of
absolute logarithmic residuals; D - Diemand et al. (2004);
R - Reed et al. (2005); 2 - $n$: total number of sample
objects; 3 - $\overline{\gamma}$: arithmetic mean; 4 -
$\sigma_{s~\gamma}$: standard deviation; 5 - $\sigma_
{s~\overline{\gamma}}$: standard deviation from the mean;
6 - $\Delta^-\gamma$: maximum negative deviation from
the mean; 7 - $\Delta^+\gamma$: maximum positive deviation
from the mean; 8 - $\Delta^\mp\gamma$: mean maximum deviation
from the mean; 9 - $\overline{\gamma}^-=\overline{\gamma}-3
\sigma_{s~\overline{\gamma}}$: lower limit assigned to the
mean; 10 - $\overline{\gamma}^+=\overline{\gamma}+
3\sigma_{s~\overline{\gamma}}$: upper limit assigned to the
mean.   Sample haloes represent clusters or groups, with
the exception of three objects belonging to the original
R sample (excluded in the reduced R sample), which
represent SDHs embedding the Milky Way and two dwarf
galaxies, respectively.}
\label{t:gamma}
\end{table}
All sample haloes are on the scale of
cluster of galaxies, with the exception
of three objects belonging to the Reed
et al. (2005) sample, which represent
SDHs embedding the Milky Way and two dwarf
galaxies, respectively.   To get a homogeneous
sample, a reduced set has been considered,
which includes SDHs embedding only clusters
of galaxies.   A marginal discrepancy between
results from the current paper and the original
Reed et al. (2005) sample is found to disappear
in dealing with the reduced sample, as shown
in Tab.\,\ref{t:gamma}.

With regard to the asymptotic inner slope
of the logarithmic density profile, the
current results conform to the validity
of the Jeans equation, which demands
$1\le\gamma\le3$ for dark matter haloes
(Hansen 2004), but shallower slopes may occur
if the effects of the baryonic component 
are considered (e.g., El-Zant et al. 2004;
Hansen 2004).   In addition, the inequality,
$\gamma<2$, related to simple analytical 
treatments of dark matter haloes (Williams
et al. 2004), is also fulfilled by the
current results, see Tabs.\,2-5.
On the contrary, the inequality, $\gamma<1$,
related to force-free halo centre and vanishing
density at infinite distance (M\"ucket \& Hoeft
2003), is only marginally consistent with the
current results.

No evident correlation is found between SDH 
dynamical state (relaxed or merging) and
asymptotic inner slope of the logarithmic
density profile, $-\gamma$, or (for SDH comparable
virial masses) scaled radius, $\xi_{{\rm vir}}$,
contrary to previous results (Ascasibar et al.
2004) related to a sample of 19 high-resolution
SDHs on the scale of both clusters of galaxies
(13 objects) and galaxies (6 objects), with regard 
to NFW density profiles.   An investigation on
richer samples could provide more information
to this respect.

The GPL density profiles which best fit
to the averaged SDH density profile, are
characterized by exponents, $(\alpha,\beta,
\gamma)$, satisfying $[\Nint(\alpha),\Nint
(\beta),$ $\Nint(\gamma)]=($1,3,1), the last
related to NFW density profile.   But the 
corresponding deviations are not negligible
with the exception of the one from $\gamma$.   
The comparison with values averaged on the 
whole halo sample, discloses that the
exponents, $\beta$ and $\gamma$, seem to
fluctuate around their NFW counterparts,
but the same does not hold for $\alpha$,
which has a mean value of about 0.6.
Accordingly, NFW density profiles cannot
be conceived as universal, in the sense
mentioned above, with regard to the 
current halo sample.

On the other hand, following e.g., Bullock
et al. (2001), NFW density profiles (or
alternative functional forms) may be
considered as a convenient way to parametrize
SDH density profiles, without implying that
it necessarily provides the best possible fit.
This is why the scaled radius, $\xi_{{\rm vir}}$,
and the scaled mass, $\nu_{mas}$, can be
interpreted as general structure parameters,
not necessarily restricted to a specific
density profile.   In particular, any spread 
in $\xi_{{\rm vir}}$ and $\nu_{mas}$ can be
attributed to a real scatter in a ``physical''
scaling radius, defined by e.g., Eq.\,(\ref
{eq:dlogf1}), rather than to inaccuracies in
the assumed, ``universal'' density profile.
For further details see e.g., Bullock et al.
(2001).

Additional support to the above considerations
is provided by the value calculated for the
standard deviation of the decimal logarithm
of the scaled radius, $\sigma_{s\log\xi_{{\rm vir}}}
=0.15$-0.17, which is very close to $\sigma_
{s\log\xi_{{\rm vir}}}=0.18$ deduced from a statistical
sample of about five thousands of simulated
haloes, within mass bins equal to (0.5-1.0)
$\times10^nh^{-1}{\rm M}_\odot$, where $11\le
n\le14$ and $n$ is an integer (Bullock et al.
2001).

The standard deviations listed in Tabs.\,\ref
{t:para} and \ref{t:parq}, related to 
characteristic parameters of best fitting GPL
density profiles, with regard to sample haloes,
have been calculated under the assumption that
they obey a Gaussian distribution, using
Eqs.\,(\ref{eq:etam})-(\ref{eq:smum}).   The
existence of a Gaussian distribution is a
necessary, but not sufficient condition, for
the validity of the central limit theorem. 
In this view, the parameters under discussion
are related to the final properties of the
corresponding sample halo, which are connected
with the initial conditions, $\alpha_1$,
$\alpha_2$, ..., $\alpha_n$, intended as 
random variables, by a transformation,
$\eta^\ast=\alpha_1\cdot\alpha_2\cdot...\cdot
\alpha_n$.   For further details, see Caimmi
\& Marmo (2004).

In addition, it is worth of note that the
application of a least-absolute values or a 
least-squares method (in particular RFSM5),
in fitting GPL to SDH density profiles (e.g.,
Dubinski \& Carlberg 1991; Klypin et al. 2001;
Fukushige \& Makino 2003) implies a (fiducial)
Gaussian distribution of the SDH density  
profile, $y_{SDH}=\log(\rho_{SDH}/\rho_h)$,
around the expected value deduced from the
related GPL density profile, $y_{GPL}=\log
(\rho_{GPL}/\rho_h)$, for any fixed logarithmic
radial bin centered on $x=\log[(r_{i+1}+r_i)
/(2r_{{\rm vir}})]$. 
It is the particularization, to the case of
interest, of a well known result of the theory 
of errors (e.g., Taylor 2000, Chap.\,8,
\S\,8.2).

The results of the current paper confirm a
certain degree of degeneracy in fitting GPL
to SDH density profiles, as pointed out by
Klypin et al. (2001).   For instance, four
GPL density profiles fit to the sample halo
$S02.10$, where the sum of absolute values and
squares of absolute logarithmic residuals,
$\sum\vert R_i\vert$ and $\sum R_i^2$, the
exponents, $\alpha$, $\beta$, and $\gamma$, 
the scaled radius, $\xi_{{\rm vir}}$, the scaled
mass, $\nu_{mas}$, the scaling radius, $r_
0$, and the scaling density, $\rho_0$, lie
within the following ranges:
\begin{leftsubeqnarray}
\slabel{eq:inta}
&& 0.675<\sum\vert R_i\vert<0.732~~; \\
\slabel{eq:intb}
&& 0.0509<\sum R_i^2<0.0537~~; \\
\slabel{eq:intc}
&& 0.559<\alpha<0.727~~; \\
\slabel{eq:intd}
&& 2.73<\beta<3.37~~; \\
\slabel{eq:inte}
&& 0.818<\gamma<0.943~~; \\
\slabel{eq:intf}
&& 3.89<\xi_{{\rm vir}}<8.04~~; \\
\slabel{eq:intg}
&& 0.431<\nu_{mas}<3.10~~; \\
\slabel{eq:inth}
&& 409<r_0/{\rm kpc}<847~~; \\
\slabel{eq:inti}
&& 1.28<10^4\rho_0/(10^{10}{\rm M}_\odot/
{\rm kpc}^3)<1.70~~;
\label{seq:int}
\end{leftsubeqnarray}
which could be explained in a twofold manner.

On one hand, a degeneracy could be intrinsic
to the 6-dimension hyperspace where the RFSM5
method works.  Accordingly, the 5-dimension
hypersurface, $w=F(x_C,y_C,b,b_\beta,b_\gamma)$,
defined by the sum of absolute values or squares
of absolute logarithmic residuals, happens to be
parallel, in some finite region of the domain, 
to the principal 5-dimension hyperplane, $({\sf 
O}~x_C~y_C~b~b_\beta~b_\gamma)$, which implies 
infinite extremum points of minimum%
\footnote{An example in a 3-dimension space,
$({\sf O}xyz)$, is provided by the surface 
of a cylinder: in the special case where
the height, or a basis, is parallel to the
principal plane, $({\sf O}xy)$, infinite
extremum points of minimum occur.   If
otherwise, there is a single extremum point 
of minimum.}.%

On the other hand, a degeneracy could be
owing to the restricted domain of SDH density
profiles, defined by Eq.\,(\ref{eq:doeta}),
which is limited by the virial radius on the 
right and by the occurrence of numerical
artifacts (mainly two-body relaxation) on
the left.   In this view, a more extended
range could reduce the degeneracy.   As the 
fit must necessarily be restricted to the 
virialized region, one shall wait for
higher-resolution simulations involving
five-parameter fits to test this
possibility.   But in recent high-resolution
simulations two or three-parameter fits
only have been used (e.g., Navarro et al.
2004; Diemand et al. 2004; Reed et al.
2005).

The current attempt is limited to GPL
density profiles, defined by Eq.\,(\ref
{eq:GPL}), but different alternatives
have been exploited in the literature,
such as the Burkert (1995) density profile:
\begin{equation}
\label{eq:prob}
\rho=\rho_0\left[1+\left(\frac r{r_0}\right)^2
\right]^{-1}\left[1+\frac r{r_0}\right]^{-1}~~;
\end{equation}
which resembles the NFW density profile for
$r\appgeq0.02r_{{\rm vir}}$.   The corresponding
scaling and scaled radii may be related as:
$(r_0)_B=(r_0)_{NFW}/1.52$; $(\xi_{{\rm vir}})_B=
1.52 (\xi_{{\rm vir}})_{NFW}$; respectively (e.g.,
Bullock et al. 2001).

Another possibility is a profile that curves
smoothly over to a constant density at very
small radii (Navarro et al. 2004):
\begin{equation}
\label{eq:expro}
\rho=\rho_0\exp\left\{-\frac2\lambda\left[
\left(\frac rr_0\right)^\lambda-1\right]
\right\}~~;
\end{equation}
where the parameter, $\lambda$, prescribes
how fast the density profile turns away from
a power-law near the centre.   In logarithmic
form, Eq.(\ref{eq:expro}) represents a class
(defined by the parameter, $-2/\lambda$) of
Sersic (1968) density profiles (Merritt et al.
2005).   The best fit
reads (19 sample objects, Navarro et al. 2004):
$\overline{\lambda}\mp3\sigma_{s~\overline
{\lambda}}=0.17216\mp0.021897$.

On the other hand, the RFSM5 method may be
extended to any kind of density profile,
keeping in mind that different classes
could exhibit different geometrical properties.

The ``universality'' of density profiles
involving scaled parameters, has to be 
intended as in polytropes (e.g., Caimmi 1980).
\begin{trivlist}
\item[\hspace\labelsep{\bf }] \sl
A single distribution in the abstract space
of the scaled variables, $\phi=f(\xi)$,
corresponds to $\infty^2$ distributions
in the physical space, $\rho/\rho_0=f(r/r_0)$,
provided the free parameters are the scaling 
radius, $r_0$, and the scaling density, $\rho
_0$, and the scaled radius reads $\xi=r/r_0$.
\end{trivlist}
In dealing with SDH density profiles, it would
be more germane to the matter speaking about
best fitting, instead of universal, density
profiles.   The validity of the fit has to be
restricted to a fiducial range where simulations
are not affected by spurious effects such as
two-body relaxation, according to e.g., Eq.\,(\ref
{eq:doeta}).   In particular, the asymptotic inner
slope is necessary for the definition of GPL density
profiles, but any conclusion outside the above
mentioned fiducial range may be at risk, in absence
of some kind of (direct or indirect) observational
support.

\section{Conclusion}\label{conc}

Analytical and geometrical properties
of generalized power-law (GPL) density profiles
have been investigated in detail.   In particular, a
one-to-one correspondence has been found between 
mathematical parameters (a scaling radius, $r_0$,
a scaling density, $\rho_0$, and three exponents,
$\alpha$, $\beta$, $\gamma$), and geometrical 
parameters (the coordinates of the intersection 
of the asymptotes, $x_C$, $y_C$, and three 
vertical intercepts, $b$, $b_\beta$, $b_\gamma$,
related to the curve and the asymptotes,
respectively): $(r_0,\rho_0,\alpha,\beta,\gamma)
\leftrightarrow(x_C,y_C,b,b_\beta,b_\gamma)$.

Then GPL density profiles have been compared with
simulated dark haloes (SDH) density profiles,
and nonlinear least-absolute values and
least-squares fits involving the above mentioned
five parameters (RFSM5 method) have been prescribed.
More specifically, the sum of absolute values or 
squares of absolute logarithmic residuals, $R_i=
\log\rho_{SPH}(r_i)-\log\rho_{GPL}(r_i)$, has been
evaluated on $10^5$ points making a 5-dimension
hypergrid, through a few iterations.   The
size has progressively been reduced around a
fiducial minimum, and superpositions on nodes
of earlier hypergrids have been avoided. 

An application has been
made to a sample of 17 SDHs on the scale of
cluster of galaxies, within a flat $\Lambda$CDM
cosmological model (Rasia et al. 2004).   In dealing
with the mean SDH density profile, a virial
radius, $r_{{\rm vir}}$, averaged over the whole
sample, has been assigned, which has allowed the calculation
of the remaining parameters.   Using a RFSM5
method has provided a better fit with respect to
other methods.   

The geometrical parameters,
averaged over the whole sample of best fitting
GPL density profiles, have yielded $(\alpha,\beta,
\gamma)\approx(0.6,3.1,1.0)$, to be compared
with $(\alpha,\beta,\gamma)=(1,3,1)$, i.e. the
NFW density profile (Navarro et al. 1995, 1996,
1997); $(\alpha,\beta,\gamma)=(1.5,3,1.5)$
(Moore et al. 1998, 1999); $(\alpha,\beta,
\gamma)=(1,2.5,1)$ (Rasia et al. 2004); and,
in addition, $\gamma\approx1.5$ (Hiotelis 2003),
deduced from the application of a RFSM5 method,
but using a different definition of scaled
radius, or concentration; $\gamma\approx$1.2\,-1.3
deduced from more recent fits (Diemand et al. 2004;
Reed et al. 2005).
No evident correlation has been found between SDH 
dynamical state (relaxed or merging) and
asymptotic inner slope of the logarithmic
density profile or (for SDH comparable
virial masses) scaled radius.

Mean values and standard
deviations of some parameters have been calculated,
and in particular the decimal logarithm of the
scaled radius, $\xi_{{\rm vir}}$, has been found to yield
$<\log\xi_{{\rm vir}}>=0.74$ and $\sigma_{s
\log\xi_{{\rm vir}}}=0.15$-0.17, consistent with previous
results related to NFW density profiles.   It
has provided additional support to the idea, that
NFW density profiles may be considered as a
convenient way to parametrize SDH density
profiles, without implying that it necessarily
produces the best possible fit (Bullock et al. 
2001).   

A certain degree of degeneracy has been
found in fitting GPL to SDH density profiles.
If it is intrinsic to the RFSM5 method or it 
could be reduced by the next generation of
high-resolution simulations, still remains
an open question.

Future work demands a generalization of the
above results on two respects.   First, the
method could be applied to data related to
dynamical mass distributions inferred in
cluster of galaxies (or galaxies), and
the fit be compared with its counterpart
deduced from simulated density profiles.
Second, the method could be applied using
Sersic (1968) density profiles in place
of the family defined by Eq.\,(\ref{eq:GPL}).
An advantage is that both baryonic and dark
(non baryonic) mass distributions are well
represented by the Sersic law (Merritt et
al. 2005), which depends on four parameters
instead of five.

\section{Acknowledgements}
We are indebted with E. Rasia, G. Tormen, and 
L. Moscardini, for making the results of their
simulations, investigated in RTM, available to us.
In addition, we
are deeply grateful to all of them for clarifying
and fruitful discussions.   We thank two anonymous
referees for useful critical comments.

\section{Appendix}

\subsection{A. Analytical and geometrical properties of
logarithmic GPL density profiles}\label{geGPL}

The values of the vertical intercept related to 
the curve, $b$, and to the asymptotes, $b_\beta$ and
$b_\gamma$, are readily determined by putting
$\log\eta=0$ i.e. $\eta=1$ in Eqs.\,(\ref
{eq:lpsi}), (\ref{eq:pas0}), and (\ref{eq:pasi}),
respectively.   The result is:
\begin{lefteqnarray}
\label{eq:b}
&& b=\log\Delta_{{\rm vir}}-\log\nu_{mas}+3\log\xi_{{\rm vir}}-
\gamma\log\xi_{{\rm vir}}-\chi\log(1+\xi_{{\rm vir}}^\alpha)~~; \\
\label{eq:bga}
&& b_\gamma=\log\Delta_{{\rm vir}}-\log\nu_{mas}+3\log\xi_{{\rm vir}}-
\gamma\log\xi_{{\rm vir}}~~; \\
\label{eq:bbe}
&& b_\beta=\log\Delta_{{\rm vir}}-\log\nu_{mas}+3\log\xi_{{\rm vir}}-
\beta\log\xi_{{\rm vir}}~~;
\end{lefteqnarray}
for sake of brevity, let us denote the 
intersection of the asymptotes in the 
logarithmic plane, $({\sf O}~\log\eta~
\log\psi)$, as ${\sf C}(x_C,y_C)$ where, 
owing to Eqs.\,(\ref{eq:rho0h}), (\ref
{eq:csiv}), (\ref{eq:pas0}), (\ref
{eq:pasi}), (\ref{eq:bga}), and (\ref
{eq:bbe}), the explicit expression of 
the coordinates reads:
\begin{lefteqnarray}
\label{eq:xC}
&& x_C=-\log\xi_{{\rm vir}}=-\log\frac{r_{{\rm vir}}}{r_0}=
\log\frac{r_0}{r_{{\rm vir}}}~~; \\
\label{eq:yC}
&& y_C=\log\frac{\Delta_{{\rm vir}}\xi_{{\rm vir}}^3}{\nu_
{mas}}=\log\frac{\rho_0}{\rho_h}~~;
\end{lefteqnarray}
which yields the following.

\begin{trivlist}
\item[\hspace\labelsep{\bf Theorem.}] \sl
For a selected (but arbitrary) SDH density
profile, cosmological model, and GPL density
profile, the intersection of the asymptotes
in the logarithmic plane, $({\sf O}~\log\eta~
\log\psi)$, occurs at a point, ${\sf C}(x_C,
y_C)$, where the coordinates are the
decimal logarithm of the ratio between 
scaling radius and virial radius, and
scaling density and mean (matter) density
of the universe, respectively.
\end{trivlist}

The combination of Eqs.\,(\ref{eq:rho0h}), 
(\ref{eq:csiv}), (\ref{eq:xC}), and (\ref
{eq:yC}) yields:
\begin{equation}
\label{eq:anxy}
\log\nu_{mas}=\log\Delta_{{\rm vir}}-3x_C-y_C~~;
\end{equation}
accordingly, the vertical intercepts of the
curve and the asymptotes, expressed by
Eqs.\,(\ref{eq:b}), (\ref{eq:bga}), and 
(\ref{eq:bbe}), reduce to:
\begin{lefteqnarray}
\label{eq:bc}
&& b=y_C+\gamma x_C-\chi\log\left[1+\exp_{10}(-\alpha
x_C)\right]~~; \\
\label{eq:bgac}
&& b_\gamma=y_C+\gamma x_C~~; \\
\label{eq:bbac}
&& b_\beta=y_C+\beta x_C~~;
\end{lefteqnarray}
where, in general, $\exp_ux=u^x$, and $\exp
x={\rm e}^x$, according to the standard
notation.

A change of coordinates, defined as:
\begin{equation}
\label{eq:xy}
x=\log\eta~~;\qquad y=\log\psi~~;
\end{equation}
makes the expressions of the curve and the
asymptotes, defined by Eqs.\,(\ref{eq:lpsi}),
(\ref{eq:pas0}), and (\ref{eq:pasi}), 
translate into: 
\begin{lefteqnarray}
\label{eq:psc}
&& y=y_C-\gamma(x-x_C)-\chi\log\{1+\exp_{10}
[\alpha(x-x_C)]\}~~; \\
\label{eq:p0c}
&& y=y_C-\gamma(x-x_C)~~;\qquad x\ll x_C~~; \\
\label{eq:pic}
&& y=y_C-\beta(x-x_C)~~;\qquad x\gg x_C~~;
\end{lefteqnarray}
with regard to the plane $({\sf O}xy)$.

\begin{figure}
\centerline{\psfig{file=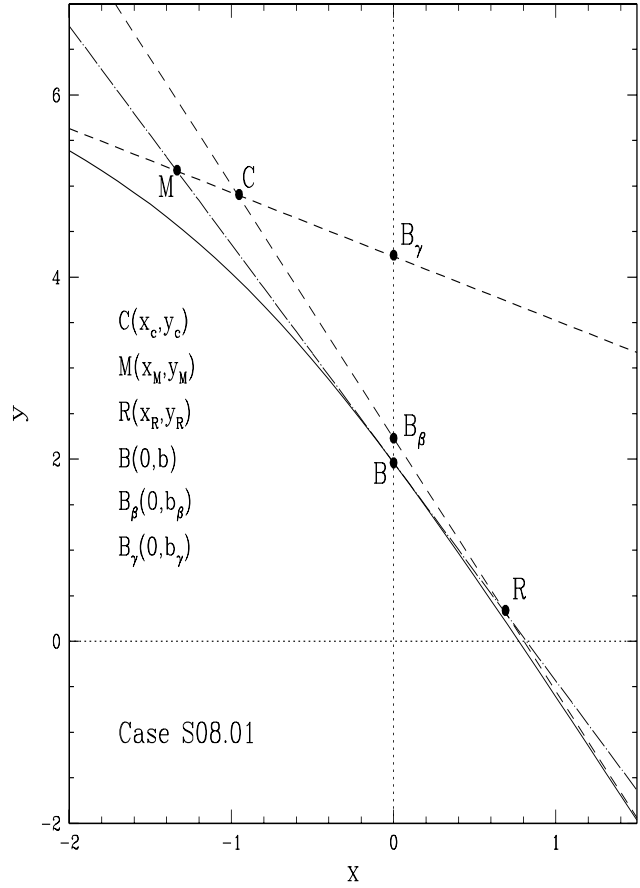,height=130mm,width=90mm}}
\caption{A NFW logarithmic density profile, together 
with the tangent at the vertical intercept and the
asymptotes, on the $({\sf O}xy)$ plane.   The above
mentioned straight lines define a triangle, {\sf CMR}.
The curve must necessarily lie below the asymptotes  
and the segment, $\overline{{\sf MR}}$, with the exception
of the tangential point, {\sf B}.}
\label{f:GPLr}
\end{figure}

The first derivative of the curve, defined by
Eq.\,(\ref{eq:psc}), takes the expression:
\begin{equation}
\label{eq:dyx}
\frac{\diff y}{\diff x}=-\beta+\frac{\beta-
\gamma}{1+\exp_{10}[\alpha(x-x_C)]}~~;
\end{equation}
and the particularization of the above result
to the vertical intercept of the curve, reads:
\begin{equation}
\label{eq:dy0}
\left(\frac{\diff y}{\diff x}\right)_{x=0}=
-\beta+\frac{\beta-\gamma}{1+\exp_{10}(-\alpha
x_C)}~~;
\end{equation}
according to Eqs.\,(\ref{eq:bc}) and (\ref
{eq:dy0}), the equation of the tangent to
the curve at the vertical intercept is:
\begin{equation}
\label{eq:tin}
y=\left[-\beta+\frac{\beta-\gamma}{1+\exp_{10}
(-\alpha x_C)}\right]x+b~~;
\end{equation}
the intersections of this line with the 
asymptotes, expressed by Eqs.\,(\ref{eq:p0c})
and (\ref{eq:pic}), let they be ${\sf M}(x_M,
y_M)$ and ${\sf R}(x_R,y_R)$, respectively,
may be calculated after some algebra.   The
result is:
\begin{lefteqnarray}
\label{eq:xM}
&& x_M=\frac{b_\beta-b}{\beta-\gamma}\left[
1+\exp_{10}(-\alpha x_C)\right]~~; \\
\label{eq:yM}
&& y_M=b_\beta-\beta\frac{b_\beta-b}{\beta-
\gamma}\left[1+\exp_{10}(-\alpha x_C)\right]~~; \\
\label{eq:xR}
&& x_R=-\frac{b_\gamma-b}{\beta-\gamma}\left[
1+\exp_{10}(-\alpha x_C)\right]~~; \\
\label{eq:yR}
&& y_R=b_\gamma+\gamma\frac{b_\gamma-b}{\beta-
\gamma}\left[1+\exp_{10}(-\alpha x_C)\right]~~;
\end{lefteqnarray}
the curve, the tangent to the curve at the
vertical intercept, and the asymptotes, are
represented in Fig.\,\ref{f:GPLr}.
It is apparent that the curve must necessarily lie 
below the asymptotes and the segment, $\overline{{\sf MR}}$, 
with the exception of the tangential point, ${\sf B}
(0,b)$.

The combination of Eqs.\,(\ref{eq:bc}), (\ref{eq:bgac}),
and (\ref{eq:bbac}) yields:
\begin{lefteqnarray}
\label{eq:bgab}
&& b_\gamma-b=\chi\log\left[1+\exp_{10}(-\alpha x_C)
\right]~~; \\
\label{eq:bgabe}
&& b_\beta-b_\gamma=(\beta-\gamma)x_C=\chi\alpha x_C~~;
\end{lefteqnarray}
and the comparison between the alternative
expressions of the exponent, $\chi$, deduced
from Eqs.\,(\ref{eq:bgab}), (\ref{eq:bgabe}),
produces:
\begin{leftsubeqnarray}
\slabel{eq:traw}
&& Aw=\log\left[1+\exp_{10}w\right] \\
\slabel{eq:A}
&& A=\frac{b_\gamma-b}{b_\gamma-b_\beta}>1~~; \\
\slabel{eq:w}
&& w=-\alpha x_C>0~~;
\label{seq:tra}
\end{leftsubeqnarray}
which is equivalent to:
\begin{leftsubeqnarray}
\slabel{eq:trau}
&& u^A-u-1=0~~; \\
\slabel{eq:u}
&& u=\exp_{10}w ~~; 
\label{seq:trau}
\end{leftsubeqnarray}
where $A>1$, $w>0$, and $-\beta<-\gamma<0$,
in the case of interest (logarithmic GPL 
density profiles of the kind represented in 
Fig.\,\ref{f:GPLr}).

The function, $\phi(u)$, on the left-hand side
of Eq.\,(\ref{eq:trau}), has the following
properties:
\begin{leftsubeqnarray}
\slabel{eq:phia}
&& \phi(u)=u^A-u-1~~;\qquad0\le u<+\infty~~; \\
\slabel{eq:phib}
&& \phi(0)=-1~~;\qquad u_{min}=A^{-1/(A-1)}~~; \\
\slabel{eq:phic}
&& \lim_{u\to+\infty}\phi(u)=+\infty\qquad
\phi(u_0)=0~~;
\label{seq:phi}
\end{leftsubeqnarray}
where $u_{min}$ and $u_0$ denote the abscissa
of the extremum point (of minimum) and the
zero of the function, respectively.   Then
Eq.\,(\ref{eq:trau}) may be solved by use
of an iterative method.

The combination of Eqs.\,(\ref{eq:bgabe}),
(\ref{eq:w}), and (\ref{eq:u}) yields:
\begin{lefteqnarray}
\label{eq:alfa}
&& \alpha=-\frac{\log u_0}{x_C}~~; \\
\label{eq:chi}
&& \chi=\frac{b_\gamma-b_\beta}{\log u_0}~~;
\end{lefteqnarray}
on the other hand, the exponents, $\gamma$
and $\beta$, may be deduced from Eqs.\,(\ref
{eq:bgac}) and (\ref{eq:bbac}), respectively,
as:
\begin{lefteqnarray}
\label{eq:gamma}
&& \gamma=\frac{b_\gamma-y_C}{x_C}~~; \\
\label{eq:beta}
&& \beta=\frac{b_\beta-y_C}{x_C}~~;
\end{lefteqnarray}
and the scaling parameters, $r_0$ and $\rho_0$,
may be deduced from Eqs.\,(\ref{eq:xC}) and 
(\ref{eq:yC}), respectively, as:
\begin{lefteqnarray}
\label{eq:r0}
&& r_0=r_{{\rm vir}}\exp_{10}x_C~~; \\
\label{eq:rho0}
&& \rho_0=\rho_h\exp_{10}y_C~~;
\end{lefteqnarray}
the set of Eqs.\,(\ref{eq:alfa})-(\ref{eq:rho0})
yields the following.
\begin{trivlist}
\item[\hspace\labelsep{\bf Theorem.}] \sl
For a selected (but arbitrary) SPH density
profile, cosmological model, and GPL density
profile, in the logarithmic plane, $({\sf O}
xy)$, there is a one-to-one correspondence
between analytical parameters, $(r_0,\rho_0,
\alpha,\beta,\gamma)$, and geometrical 
parameters, $(x_C,y_C,b,b_\beta,b_\gamma)$,
in the sense that either set univocally
determines a GPL density profile.
\end{trivlist}
The advantage of using geometrical instead
of analytical parameters lies in a better
understanding of how the curve change as
one or more parameters do. 

With regard to SDH density profiles, 
according to Eqs.\,(\ref{eq:doeta}) and
(\ref{eq:xy}), let us divide the domain
into two distinct regions, as:
\begin{equation}
\label{eq:xgb}
-2\le x_\gamma\le-1~~;\qquad-1\le x_\beta\le0~~;
\end{equation}
which shall be called the $\gamma$ region
and the $\beta$ region, respectively.

Let $(x_i,y_i)$ be coordinates of a generic
point of a logarithmic SDH density profile, 
and $[x_i,y(x_i)]$ their counterparts related 
to a fitting, logarithmic GPL density
profile.   Owing to Eq.\,(\ref{eq:psc}), the
corresponding, logarithmic absolute residual,
is:
\begin{lefteqnarray}
\label{eq:Ri}
&& R_i=y_i-y(x_i)=y_i-y_C+\gamma(x_i-x_C) \nonumber \\
&& \phantom{R_i=}+\chi\log\{1+\exp_{10}[\alpha(x_i-x_C)
]\}~~;
\end{lefteqnarray}
the particularization of Eq.\,(\ref{eq:Ri}) to 
the $\gamma$ and $\beta$ region, defined by
Eq.\,(\ref{eq:xgb}), allows the application
of a least-squares fit to the related portions
of SDH density profile.   The best linear
fites are:
\begin{lefteqnarray}
\label{eq:mqgs}
&& y=b_{s\gamma}-\gamma_sx~~; \\
\label{eq:mqbs}
&& y=b_{s\beta}-\beta_sx~~;
\end{lefteqnarray}
and the coordinates of their intersection
point, ${\sf C}_s(x_s,y_s)$, are:
\begin{equation}
\label{eq:ints}
x_s=\frac{b_{s\gamma}-b_{s\beta}}{\beta_s-\gamma_s}
~~;\qquad y_s=\frac{b_{s\gamma}\beta_s-b_{s\beta}
\gamma_s}{\beta_s-\gamma_s}~~;
\end{equation}
the best linear fits to a selected
SDH density profile, are plotted in Fig.\,\ref
{f:SDH}.

The intercepts, $b_{s\gamma}$ and $b_{s\beta}$,
and the slopes, $-\gamma_s$ and $-\beta_s$, 
appearing in Eqs.\,(\ref{eq:mqgs}) and (\ref
{eq:mqbs}), are calculated using the standard
formulation of the method (e.g., Secco 2001,
Chap.\,4, \S\,4.1):
\begin{lefteqnarray}
\label{eq:camq}
&& -\lambda_s=\frac{\overline{x_\lambda y_\lambda}-
\overline{x}_\lambda\overline{y}_\lambda}{\overline{x_\lambda 
x_\lambda}-\overline{x}_\lambda\overline{x}_\lambda}~~; \\
\label{eq:inmq}
&& b_{s\lambda}=\frac{\overline{y}_\lambda\overline{x_
\lambda x_\lambda}-\overline{x}_\lambda\overline{x_\lambda
x_\lambda}}{\overline{x_\lambda 
x_\lambda}-\overline{x}_\lambda\overline{x}_\lambda}~~; \\
\label{eq:vcamq}
&& \sigma_{s~\lambda_s}^2=\sigma_{s~y_\lambda}^2\frac
1{\overline{x_\lambda x_\lambda}-\overline{x}_\lambda\overline{x}_
\lambda}~~; \\
\label{eq:vinmq}
&& \sigma_{s~b_s}^2=\sigma_{s~y_\lambda}^2\frac
{\overline{x_\lambda x_\lambda}}{\overline{x_\lambda 
x_\lambda}-\overline{x}_\lambda\overline{x}_\lambda}~~;
\end{lefteqnarray}
where $\lambda=\beta,\gamma$, a bar denotes
arithmetic mean over the corresponding range
of simulated values, and $\sigma_s^2$ is the
empirical variance of the histogram calculated
for the selected random variable.

The comparison between best linear fits
to SDH density profiles, defined by Eqs.\,(\ref
{eq:mqgs}) and (\ref{eq:mqbs}), and related asymptotes
of GPL density profiles, defined by Eqs.\,(\ref
{eq:p0c}) and (\ref{eq:pic}), implies the
following, fiducial conclusions: (i) SDH best
linear fits lie below related GPL asymptotes,
in the range of interest, and (ii) SDH best 
linear fits are more inclined (in absolute
value) with respect to related GPL asymptotes
towards negative infinite, and less inclined
(in absolute value) with respect to related
GPL asymptotes towards positive infinite.
Accordingly (iii) the intersection between SDH
best linear fits lies below, and on the left,
with respect to the intersection of related GPL
asymptotes.

The above mentioned conclusions (i) and (ii)
read:
\begin{lefteqnarray}
\label{eq:bsb}
&& b<b_{s\beta}~~;\qquad b_{s\beta}<b_\beta~~;
\qquad b_{s\gamma}<b_\gamma~~; \\
\label{eq:begas}
&& \gamma_s<\gamma<0~~;\qquad\beta<\beta_s<
\gamma_s~~;
\end{lefteqnarray}
on the other hand, GPL asymptotes intersect
within the $(-+)$ quadrant, according to
Eqs.\,(\ref{eq:xC}) and (\ref{eq:yC}), which
yields:
\begin{equation}
\label{eq:beby}
b<y_C~~;\qquad b_\beta<y_C~~;\qquad b_\gamma<
y_C~~;
\end{equation}
finally, lower and upper values to the ranges:
\begin{equation}
\label{eq:xyc}
y_{min}<y_C<y_{max}~~;\qquad x_{min}<x_C<x_{max}~~;
\end{equation}
and the lower value to the range:
\begin{equation}
\label{eq:bmin}
b_{min}<b<b_{s\beta}~~;
\end{equation}
may be deduced from SDH density profiles.
The upper value of inequality (\ref{eq:bmin})
is owing to the first inequality (\ref{eq:bsb}). 
The combination of inequalities (\ref{eq:bsb}) 
and (\ref{eq:beby}) yields:
\begin{equation}
\label{eq:bsby}
b_{s\beta}<b_\beta<y_C~~;\qquad b_{s\gamma}<b_
\gamma<y_C~~;
\end{equation}
due to negative slopes of GPL asymptotes,
$-\gamma<0$ and $-\beta<0$. 

With regard to a 5-dimension hyperspace,
$({\sf O}~x_C~y_C~b~b_\beta~b_\gamma)$, inequalities
(\ref{eq:xyc}), (\ref{eq:bmin}), and (\ref
{eq:bsby}), define a 5-dimension hyperrectangle 
of sides $(x_{max}-x_{min})$, $(y_{max}-y_{min})$, 
$(b_{s\beta}-b_{min})$, $(y_{max}-b_{s\beta})$,
and $(y_{max}-b_{s\gamma})$, respectively.
Then it is possible to make a 5-dimension
hypergrid where the points are equally spaced;
according to inequalities (\ref{eq:bsby}),
the ranges $y_C<b_\beta<y_{max}$ and $y_C<
b_\gamma<y_{max}$ are to be excluded as
outside the cases of interest.   In dealing
with the remaining points, the sum of both
absolute values and squares of absolute
logarithmic residuals, defined by Eq.\,(\ref
{eq:Ri}), is performed, and the minimum on
the hypergrid is determined in each alternative.
Let it be $\left(x_C^{(a)},y_C^{(a}),b^{(a)},b_
\beta^{(a)},b_\gamma^{(a)}\right)$ and $\left(x_C^{(s)},
y_C^{(s}),b^{(s)},b_\beta^{(s)},b_\gamma^{(s)}\right)$, 
where the indices, $a$ and $s$, denote absolute
value and square, respectively.   

The next
iteration is in connection with a 5-dimension
hyperrectangle, which has the following features:
(1) it is centered near the previously determined
point of minimum; (2) it is reduced in size by a
factor of about 10, provided inequalities (\ref
{eq:xyc}), (\ref{eq:bmin}), and (\ref{eq:bsby}) 
continue to hold; and (3) there is no point in
common with the earlier hypergrid.
Then two additional
points of minimum are calculated and the
next iteration is allowed to start.

The computations end when the sum of absolute
values and squares of absolute logarithmic
residuals fall below an assigned treshold,
which yields:
\begin{lefteqnarray}
\label{eq:Rifa}
&& \sum_{i=1}^N\vert R_i\vert<\epsilon^{(a)}~~; \\
\label{eq:Rifs}
&& \sum_{i=1}^NR_i^2<\epsilon^{(s)}~~;
\end{lefteqnarray}
where the sum is performed on the range of 
interest, defined by Eq.\,(\ref{eq:xgb}).
\end{document}